\documentclass[aps,prd,reprint,superscriptaddress,amsmath,amssymb,amsfonts,showpacs]{revtex4-1}

\usepackage{graphicx}
\usepackage{bm}
\usepackage{calc}
\usepackage{color}
\usepackage{wrapfig}
\usepackage{url}
\usepackage{paralist}
\usepackage{array}
\usepackage{tabularx}

\usepackage{txfonts}
\DeclareMathOperator \arcsinh {arsinh}

\def\araa{ARA\&A}%
\def\apj{ApJ}%
\def\apjl{ApJ}%
\def\apjs{ApJS}%
\def\ao{Appl.~Opt.}%
\def\aap{A\&A}%
\def\aaps{A\&AS}%
\def\mnras{MNRAS}%
\def\prd{Phys.~Rev.~D}%
\def\pasp{PASP}%
\def\nat{Nature}%
\def\physrep{Phys.~Rep.}%
\begin{document}


\title{Substructure lensing in galaxy clusters as a constraint on low-mass sterile neutrinos\protect\\ in tensor-vector-scalar theory: The straight arc of Abell 2390}


\author{Martin Feix}
\email[Electronic address: ]{mf256@st-andrews.ac.uk}
\affiliation{SUPA, School of Physics and Astronomy, University of St Andrews, North Haugh, KY16 9SS, St Andrews, United Kingdom}
\author{HongSheng Zhao}
\affiliation{SUPA, School of Physics and Astronomy, University of St Andrews, North Haugh, KY16 9SS, St Andrews, United Kingdom}
\affiliation{Leiden Observatory, Leiden University, Niels Bohrweg 2, 2333 CA Leiden, The Netherlands}
\author{Cosimo Fedeli}
\affiliation{Dipartimento di Astronomia, Universit\a`{a} di Bologna, Via Ranzani 1, 40127 Bologna, Italy}
\affiliation{INFN, Sezione di Bologna, Viale Berti Pichat 6/2, 40127 Bologna, Italy}
\author{Jos\'{e} Luis Garrido Pesta\~{n}a}
\affiliation{Departamento de F\'{i}sica, Universidad de Ja\'{e}n, Campus Las Lagunillas, 23071 Ja\'{e}n, Spain}
\author{Henk Hoekstra}
\affiliation{Leiden Observatory, Leiden University, Niels Bohrweg 2, 2333 CA Leiden, The Netherlands}

\date{\today}

\begin{abstract}

Certain covariant theories of the modified Newtonian dynamics paradigm seem to require an additional hot dark matter (HDM) component - in the form of either heavy ordinary neutrinos or more recently light sterile neutrinos (SNs) with a mass around $11$eV - to be relieved of problems ranging from cosmological scales down to intermediate ones relevant for galaxy clusters. Here we suggest using gravitational lensing by galaxy
clusters to test such a marriage of neutrino HDM and modified gravity, adopting the framework of tensor-vector-scalar theory (TeVeS). Unlike
conventional cold dark matter (CDM), such HDM is subject to strong phase-space constraints, which allows one to check cluster lens models inferred within the modified framework for consistency. Since the considered HDM particles cannot collapse into arbitrarily dense clumps and only form structures well above the galactic scale, systems which indicate the need for dark substructure are of particular interest. As a first example, we study the cluster lens Abell $2390$ and its impressive straight arc with the help of numerical simulations. Based on our results, we outline a general and systematic approach to model cluster lenses in TeVeS which significantly reduces the calculation complexity. We further consider a simple bimodal lens configuration, capable of producing the straight arc, to demonstrate our approach. We find that such a model is marginally consistent with the hypothesis of $11$eV SNs. Future work including more detailed and realistic lens models may further constrain the necessary SN distribution and help to conclusively assess this point. Cluster lenses could therefore provide an interesting discriminator between CDM and such modified gravity scenarios supplemented by SNs or other choices of HDM.
\end{abstract}


\maketitle


\section{Introduction}
\label{section1}
The current concordance model of cosmology has been remarkably successful in forming a consistent picture on the largest physical scales. For instance, it provides suitable explanations for the observations on supernovae Ia \cite{sne}, large-scale structure \cite{largescale,2dFGRS}, weak lensing \cite{Benjamin2007}, and the cosmic microwave background (CMB) \cite{cmb,cmb2}. Based on general relativity (GR), this so-called $\Lambda$CDM model relies on two phenomenologically motivated constituents, cold dark matter (CDM) and dark energy (DE), which account for approximately $95$\% of the energy-matter content of the Universe. From a more fundamental point of view, however, the introduction of a DE component is theoretically challenging and extremely fine-tuned, despite the many proposals for its dynamics \cite{Copeland2006}. On the other hand, the concept of CDM also suffers from several issues such as the lack of direct experimental detection \cite{Bertone2005}, the question of its cosmological abundance \cite{Bertone2010}, and problems related to the formation of structure on small scales \cite{deBlok2010,Klypin1999,Moore1999,Sellwood2001}.

A perhaps more natural solution might be that gravity genuinely differs from GR, which expresses itself as either one or even both of the above dark components \cite{Capozziello2003,Carroll2004,Dvali2000,Dvali2001,Li2009,vectornew}. Here we want to consider a particular class of modifications which give rise to what has become known as the modified Newtonian dynamics (MOND) paradigm \citep{Mond1,Mond3}. On a nonrelativistic level, MOND aims at solving the missing mass problem by postulating an acceleration-dependent change of Newton's law which is characterized by a scale $a_{0}$ \citep{mondnew}:
\begin{equation}
\tilde{\mu}\left(\frac{|\mathbf{a}|}{a_{0}}\right)\mathbf{a} = - \bm{\nabla}\Phi_{N} +\mathbf{S}.
\label{eq:1}
\end{equation}
Here, $\Phi_{N}$ denotes the common Newtonian potential of a matter source and $\mathbf{S}$ is a solenoidal vector field determined by the condition that $\mathbf{a}$ can be expressed as the gradient of a scalar potential. The function $\tilde\mu$, controlling the modification of Newton's law, has the following asymptotic behavior:
\begin{equation}
\begin{split}
\tilde\mu(x) \sim x  \qquad x \ll 1,\\
\tilde\mu(x) \sim 1  \qquad x \gg 1.
\end{split}
\label{eq:2}
\end{equation}
Equation \eqref{eq:1} has been constructed to agree with the fact that the rotation curves of spiral galaxies become flat outside their central parts. Analyzing observational data, Milgrom estimated an acceleration scale of $a_{0} \approx 1.2 \times 10^{-10}$m s$^{-2}$.

The MOND paradigm still appears suitable to explain the observed ``conspiracy'' between the distribution of baryons and the gravitational field in spiral galaxies \cite{insight2,insight}. It is striking that such a simple prescription leads to extremely successful predictions for galaxies ranging over five decades in mass (see, e.g., Refs. \cite{spiral1,mondref1} for reviews), including our own Milky Way \cite{milky,escape,mondref6}, dwarf spheroidals \cite{dwarf1,dwarf2,dwarf3}, x-ray dim elliptical galaxies \cite{mondref4,velocity1}, and tidal dwarf galaxies \cite{tidal1,tidal2}. In addition, MOND successfully reproduces empirical galaxy scaling relations such as the well-known Tully-Fisher relation \cite{McGaugh2000,McGaugh2005}, and more recently the central surface brightness predicted by dark halos \cite{Gentile2009,Milgrom2009,Donato2009}.

While the framework of MONDian dynamics works extremely well on galactic scales, the situation in galaxy groups and clusters is quite different: Several studies of such systems \cite{sanders99,sanders03,group} have shown that an additional nonluminous matter component is required to explain observations, even after taking into account the gravitational boost induced by the MOND formula. Assuming that MOND is a viable description for such gravitating systems, this result has led to the question of what the needed matter component should be. It is obvious that any possible form of exotic CDM is disfavored as it would cause the original idea of Eq. \eqref{eq:1} to become redundant.

Another problem arises from the fact that the original MOND formulation does not specify cosmology or the nature of gravitational light deflection. Recent developments in the theory of gravity, however, have been able to embed MONDian dynamics into fully Lorentz-covariant theories by means of a dynamical four-vector field \cite{teves,tv2,fieldtheo,covariant1,covariant2}. Although still lacking a derivation from fundamental principles underpinning the MOND paradigm, these theories allow for new predictions regarding cosmology and structure formation \cite{tevesneutrinocosmo,dod1,dod2,dod3} as well as gravitational lensing \cite{chiu,qin,lenstest,tevesfit,chiba,asymmetric,filamentlens}. An appealing feature of such modifications is also that they might be able to simultaneously explain the observed effects of DE \cite{Li2009,Hao2009,Bourliot2007,Diaz2006,tevescosmo,Armendariz2009,Cardone2009}, but we do not consider this possibility in the present work.

Adopting the framework of tensor-vector-scalar theory (TeVeS) \cite{teves}, it was possible to investigate the resulting CMB and matter power spectra \cite{tevesneutrinocosmo}. It has been found that the known baryonic matter content of the universe is not enough to reconcile the theory with observations \footnote{The main difficulty is to achieve an angular-distance relation which is able to match the observed position of the peaks in the angular power spectrum of the CMB.}. To overcome this problem, one may resort to a model with ordinary neutrinos of mass around $2$eV which is able to describe the observational data in a qualitatively acceptable way, but the corresponding fits do not match the excellent agreement of a $\Lambda$CDM model. Interestingly, the idea of massive neutrinos has already been discussed to provide a solution to the lack of matter on cluster scales \cite{sanders99,sanders03}, and to explain the observed weak lensing map of the galaxy merger $1$E$0657-558$ (``bullet cluster'') \cite{bullet,tevesfit}. It should be mentioned that the needed neutrino mass of $2$eV is barely consistent with the current upper limit on the electron neutrino's mass (e.g., see Ref. \cite{Kraus2004}); future measurements such as the Karlsruhe Tritium Neutrino Mass Experiment (KATRIN) \cite{katrin,katrin2} will be able to explore a mass range well below the $2$eV threshold.

Nevertheless, the rather unsatisfactory results of this solution on large scales, especially for the CMB anisotropy power spectrum, and problems within galaxy groups \cite{group} have led to deem the hypothesis of very massive ordinary neutrinos unattractive. Alternatively, the required additional matter could be provided in the form of heavy (right-handed) sterile neutrinos (SNs) which are motivated by theoretical considerations in particle physics (e.g., see Refs. \cite{Marcolli2009,Kusenko2009,Boyarsky2009} and references therein) and offer an elegant way to explain the small masses of active neutrinos via the ``seesaw mechanism'' \cite{GELL1979,YAN1980,Lindner2002}. The conceptual advantage of such an approach lies in combining the success of modified gravity on small scales with new physics in a sector of the standard model which is known to be incomplete \cite{Gonzalez2003} and in need of revision \footnote{Without resorting to a modification of gravity, SNs in the keV mass range still provide a viable candidate for all the dark matter in the universe \cite{Boyarsky2009}. In this case, however, one may expect similar fine-tuning issues on small scales as in current CDM models.}. Motivated by a possible interpretation of the MiniBooNE experiment \cite{Giunti2008}, Angus \cite{angussn} has suggested to use a single light species of SNs with a mass of approximately $11$eV and investigated its consequences. If such SNs decouple while they are relativistic and in thermal equilibrium, one should obtain both a background evolution and a CMB power spectrum which are basically indistinguishable from a standard $\Lambda$CDM cosmology \footnote{Although this has not been explicitly calculated, one can use the following argument: For common choices of the TeVeS parameters, the impact of perturbations due to the extra fields is small at early times, i.e. those relevant for the CMB. Thus the theory exhibits a GR-like behavior, which allows to directly adopt the results of Angus for TeVeS. This is further supported by the nearly identical results for the CMB power spectrum in TeVeS \cite{tevesneutrinocosmo} and GR \cite{angussn}, assuming three active neutrinos with a mass around $2$eV. However, it is still an open question whether secondary anisotropies such as the thermal or kinetic Sunyaev-Zel'dovich effects leave a different signature than in $\Lambda$CDM.}, while at the same time, this additional hot dark matter (HDM) component may give rise to a correct prediction of the linear matter power spectrum and represents a suitable candidate for the missing mass in galaxy clusters without spoiling MONDian dynamics on galactic scales \cite{sncluster}. As for the nonlinear regime of structure formation, the situation is still unclear. Because of the more sophisticated mathematical structure of the nonlinear TeVeS field equations (or that of related theories) as opposed to those of GR, there seems currently no way to gain reliable information about the nonlinear evolution. This difficulty is somewhat reflected by the fact that the resulting field equations in the quasistatic, nonrelativistic limit typically remain highly nonlinear. Assuming an {\it ad hoc} modification of the original MOND formula Eq. \eqref{eq:1}, however, a first simplified attempt into this direction is discussed in Ref. \cite{Llinares2008}.

It is noteworthy that TeVeS or TeVeS-like theories in combination with sufficiently abundant massive neutrinos provide the most consistent
relativistic MOND framework presented in the literature so far \footnote{Note that there are certain theories which aim at reproducing MOND and large-scale observations without any additional dark matter \cite{Li2009}, but it is currently unknown whether such models naturally give rise to the observed properties of galaxy clusters.}; nevertheless, there are still innumerable aspects which need to be tested further. A powerful astronomical tool to challenge these theories is gravitational lensing. Using spherically symmetric lens models, TeVeS has been tested against the CfA-Arizona Space Telescope Lens Survey data of image-splitting lens galaxies \cite{lenstest}, building on earlier work in the context of MOND \cite{qin} (also see Refs. \cite{extralenstest,extralenstest2}). Later, the analysis has been extended to nonspherical lens geometries \cite{hkmodel}. While there is currently disagreement on how well TeVeS explains the lensing properties of relatively isolated galaxies \cite{extralenstest,extralenstest2,Chiu2010}, the models consistently seem to fail for lens galaxies embedded into dense environments like groups or clusters. A detailed individual discussion on such lens systems can be found in Ref. \cite{hkmodel}. Several authors \cite{lenstest,hkmodel} have argued that this issue might be attributed to the already known dynamical problem in groups and clusters and therefore resolved if the actual environment of lens galaxies is properly taken into account. Another way of testing the theory is offered by weak galaxy-galaxy lensing. Using data from the Red-Sequence Cluster Survey
and the Sloan Digital Sky Survey (SDSS), it has been found that the most luminous galaxies ($\gtrsim 10^{11}L_{\odot}$) would require a substantial fraction of nonbaryonic matter \cite{Tian2009}. Although this result needs to be confirmed by larger data sets before a firm conclusion can be drawn, it might hint towards a problem with the original MOND idea on galactic scales. Again, SNs with a mass around $11$eV could provide a remedy as they should be able to cluster densely enough in such massive systems \cite{sncluster}. However, it remains to be seen whether such an approach can explain observations.

In this work, we suggest to test TeVeS and the massive SN hypothesis in the context of complex lens systems which are typically present in the central regions of galaxy clusters. A previous analysis \cite{Zhaoclusterlens} already revealed that such an environment can put stringent constraints on the distribution and plausibility of the needed dark neutrino component, thus providing an excellent testbed for our purposes. Generally, the advantage of galaxy clusters lies in the independent estimates of baryonic matter, inferred from observed X-ray and stellar luminosities, and of the system's total mass distribution based on a combination of weak and strong gravitational lensing. Being insensitive to the dynamical state of the deflecting mass, the latter techniques are particularly suited to constrain the properties of the dark component. In contrast to weak lensing estimates, strong lensing is basically free of statistical uncertainties and offers a unique and robust probe of the matter distribution on scales $\lesssim100$kpc. Here we shall use strong lensing to further test the viability of $11$eV SNs.
Unlike conventional CDM, light SNs are subject to strong phase-space bounds set by the Tremaine-Gunn limit \cite{TG1979}, which allows one to check cluster lens models inferred within the modified framework for consistency. Since this limit prevents SNs from clustering into dense clumps, galaxy cluster lenses with a considerable amount of dark substructure provide an ideal target for our intentions. As a first example, we shall study the galaxy cluster Abell 2390 (A2390) with its notorious straight arc, and investigate whether it is possible to reproduce this particular lens feature in TeVeS.

The paper is structured as follows: Starting with a brief introduction to TeVeS and the formalism of gravitational lensing in Sec. \ref{section2}, we give an observational summary of the galaxy cluster A2390 and its pronounced straight arc in Sec. \ref{section3}. Continuing with the setup for a simplified density model of A2390 in Sec. \ref{section4}, we discuss results for quasiequilibrium configurations in Sec. \ref{section52}. Based on the latter, we outline a systematic approach to cluster lenses in TeVeS, and describe a lens model for the straight arc in Sec. \ref{section53}. Finally, we conclude in Sec. \ref{section6}. For clarity, technical and numerical details are given in an appendix.


\section{TeVeS and gravitational lensing\protect\\ in a nutshell}
\label{section2}
In the following, we will briefly review TeVeS and gravitational lensing within the approximations used for quasistatic systems like galaxies
and galaxy clusters. Henceforth, we shall use units with $c = 1$, and the ``square'' of a vector denotes the square of its Euclidean
norm, i.e. $\mathbf{A}^{2} = {\lVert\mathbf{A}\rVert_{2}}^2$.

\subsection{The nonrelativistic limit of TeVeS}
\label{section21}
TeVeS theory \citep{teves} is a fully relativistic description of gravity which provides a covariant framework for the MOND paradigm. Compared to GR, TeVeS is based on the well-known Einstein metric $g_{\mu\nu}$ and two additional dynamical (gravitational) fields, a four-vector $U_{\mu}$ and a scalar $\phi$; it also invokes a second ``physical'' metric $\tilde g_{\mu\nu}$, related to $g_{\mu\nu}$ through a nonconformal transformation involving the new fields, and being exclusively used for the coupling of matter fields ($\mu,\nu = 0,...,3$):
\begin{equation}
\tilde g_{\mu\nu} = e^{-2\phi}g_{\mu\nu}-2U_{\mu}U_{\nu}\sinh(2\phi).
\label{eq:3}
\end{equation}
Furthermore, the theory introduces a new set of free parameters: The constants $k$ and $K$ as the couplings of the scalar and vector field, respectively, and a length scale $l$. Albeit not a unique extension, TeVeS is the most popular ``MONDian representative'' so far, and a variety of its aspects have been extensively studied in the literature (see Ref. \cite{tevesreview} for a review). Although the original formulation of TeVeS suffers from several problems, e.g. in the strong gravity regime \citep{contaldi,Sagi2009} or - at least for certain models - in the cosmological domain \cite{Reyes2010,dod3}, it still provides a viable description of relativistic MOND on extragalactic scales.

In what follows, we shall restrict ourselves to weak fields and quasistatic systems \footnote{Note a caveat here: The present approximation
ignores possible contributions arising from perturbations of the vector field $U_{\mu}$ which could have a significant impact on cluster
scales. This issue is further discussed in Sec. \ref{section6}.}.
Under these premises, the usual metric potential $\Phi$ approximately takes the form
\begin{equation}
\Phi \approx \Phi_{N} + \phi,
\label{eq:4}
\end{equation}
where the TeVeS scalar field $\phi$ obeys
\begin{equation}
\bm{\nabla}\cdot\left\lbrack\mu\left(kl^{2}\left (\bm{\nabla}\phi \right )^{2}\right)\bm{\nabla}\phi\right\rbrack = kG\rho,
\label{eq:5}
\end{equation}
and $\Phi_{N}$ corresponds to the common Newtonian potential given by Poisson's equation.
In order to reasonably recover the dynamics of MOND, the constants $k$ and $K$ have to be much less than unity, i.e. $k,K \ll 1$, and $G\approx G_{N}$, but one should also see Ref. \cite{teves} for a discussion on lower bounds of $k$. The function $\mu(y)$, which is related to a potential-like term in the scalar field's action, has to satisfy the following conditions:
\begin{equation}
\begin{split}
\mu(y) \rightarrow 1,\quad y\rightarrow\infty,\\
{
\mu(y) \sim \sqrt{\frac{y}{b}},\quad y\ll 1,
}
\label{eq:6}
\end{split}
\end{equation}
where
\begin{equation}
y = kl^{2}\left (\bm{\nabla}\phi\right )^{2},
\label{eq:6a}
\end{equation}
and $b$ is a positive real constant. The TeVeS length scale $l$ can be related to Milgrom's $a_{0}$ by
\begin{equation}
a_{0} = \frac{\sqrt{bk}}{4\pi\Xi l} \approx \frac{\sqrt{bk}}{4\pi l},
\label{eq:7}
\end{equation}
where $\Xi=1-K/2-2\phi_{c}$, and $\phi_{c}$ is the cosmological value of the scalar field which is assumed to be small ($\phi_{c} \ll 1$) \cite{teves}. If not specified in any other way, we shall work with
\begin{equation}
b = 1,\quad k = 0.01
\end{equation}
throughout this paper. This is justified following the analysis of Ref. \cite{asymmetric}: While the value of $b$ is absorbed into the definition of the scale $a_{0}$ in Eq. \eqref{eq:7}, the TeVeS lensing maps have been shown to be generally insensitive to variations of the parameter $k$ as long as it is small, $k\lesssim 0.01$.
A particular choice for the free function $\mu(y)$, suitable for the gravitational lensing analysis of A2390, will be given and discussed in Sec. \ref{section4}.

\subsection{Background cosmology}
\label{section22}
Assuming the basic fields to partake of the symmetries of the Friedmann-Robertson-Walker (FRW) spacetime, the TeVeS analog of Friedmann's equation in the Einstein frame, delineated by the metric $g_{\mu\nu}$, reads
\begin{equation}
\left(\frac{\dot a}{a}\right)^{2} = \frac{8\pi G}{3}(\rho e^{-2\phi}+\rho_{\phi})-\frac{K}{a^{2}}+\frac{\Lambda}{3},
\label{eq:8}
\end{equation}
where we have included a cosmological constant $\Lambda$, and $\rho_{\phi}$ is the energy density of the scalar field.
Using Eq. \eqref{eq:3} for $\tilde{g}_{\mu\nu}$, we switch to the matter frame and finally obtain
\begin{equation}
\frac{1}{\tilde{a}}\frac{d\tilde{a}}{d\tilde{t}} = e^{-\phi}\left(\frac{\dot{a}}{a}-\dot\phi\right),
\label{eq:9}
\end{equation}
with
\begin{equation}
d\tilde{t} = e^{\phi}dt,\quad \tilde{a} = e^{-\phi}a.
\label{eq:10}
\end{equation}
According to previous studies \citep{teves,tevesneutrinocosmo}, it is consistent to assume that the cosmological scalar field evolves slowly in time throughout cosmological history. Thus, its contribution to the Hubble expansion is negligibly small, with a ratio $\mathcal{O}(k)$ compared to the matter contribution. Setting $\rho_{\phi}=0$ and recalling that $\phi\ll 1$ at the background level, the physical Hubble parameter can be expressed as
\begin{equation}
\tilde{H}^{2} \approx H^{2} \approx H_{0}^{2}\left(\Omega_{m}(1+z)^{3}+\Omega_{\Lambda}+\Omega_{K}(1+z)^{2}\right),
\label{eq:11}
\end{equation}
where $\Omega_{K} = 1-\Omega_{\Lambda}-\Omega_{m}$ for a flat universe.

To calculate angular diameter distances in the context of gravitational lensing, we further need to specify a particular choice of parameters.
Based on the assumption of a single species of $11$eV SNs \citep{angussn,sncluster}, we shall use a flat cosmology with
$\Omega_{m} = \Omega_{b} + \Omega_{\nu} = 0.29$ and $h=0.7$, where the Hubble constant is $H_{0} = 100h$km s$^{-1}$Mpc$^{-1}$. Note that this gives a background evolution which is virtually indistinguishable from a standard $\Lambda$CDM model.

\subsection{Light bending in TeVeS and lensing formalism}
\label{section23}
Being a metric theory, the propagation of light rays in TeVeS is determined by the null geodesics of the physical metric $\tilde{g}_{\mu\nu}$. If the metric potential $\Phi$ given by Eq. \eqref{eq:4} and the peculiar velocity $v$ of the lens are small ($\Phi,v \ll 1$), one can presume a locally flat spacetime being disturbed by the potential $\Phi$; these conditions are well satisfied for galaxies and galaxy clusters.
Following the lines of Ref. \cite{gl}, the deflection angle of a light ray under these assumptions can be expressed as
\begin{equation}
\hat{\bm{\alpha}} = 2\int_{-\infty}^{\infty}\bm{\nabla}_{\bot}\Phi dl = \hat{\bm{\alpha}}_{\rm GR}+2\int_{-\infty}^{\infty}\bm{\nabla}_{\bot}\phi dl,
\label{eq:12}
\end{equation}
where $\bm{\nabla}_{\bot}$ denotes the two-dimensional gradient operator perpendicular to light propagation, and integration is performed along the unperturbed light path (Born's approximation). In addition to the deflection angle caused by the Newtonian potential $\Phi_{N}$, there is a contribution arising from the scalar field $\phi$. Assuming that the metric potential $\Phi$ is known from solving the corresponding field equations, one can directly proceed to calculate the usual lensing quantities, fully adopting the standard GR formalism \cite{teves, asymmetric} which is briefly reviewed below.

In gravitational lensing, it is convenient to introduce the deflection potential $\Psi(\bm{\theta})$ \cite{gl}:
\begin{equation}
\Psi(\bm{\theta}) = 2\frac{D_{ds}}{D_{s}D_{d}}\int\Phi(D_{d}\bm{\theta},z)dz,
\label{eq:13}
\end{equation}
where we have used $\bm{\theta}=\bm{\xi}/D_{d}$ and have chosen coordinates such that unperturbed light rays propagate parallel to the $z$ axis. Here $\bm{\xi}$ is the two-dimensional position vector in the lens plane, and $D_{s}$, $D_{d}$, and $D_{ds}$ are the (angular diameter) distances between source and observer, lens and observer, and lens and source, respectively. If a source is much smaller than the angular scale on which the lens properties change, the lens mapping can locally be linearized. Thus, the distortion of an image can be described by the Jacobian matrix
\begin{equation}
\mathcal{A}(\bm{\theta}) = \frac{\partial\bm{\beta}}{\partial\bm{\theta}} =
\begin{pmatrix}
1-\kappa-\gamma_{1} & -\gamma_{2}\\
-\gamma_{2} & 1-\kappa+\gamma_{1}
\end{pmatrix},
\label{eq:14}
\end{equation}
where $\bm{\beta}$=$\bm{\eta}/D_{s}$ and $\bm{\eta}$ denotes the two-dimensional position of the source. The convergence $\kappa$ is directly related to the deflection potential $\Psi$ through
\begin{equation}
\kappa = \frac{1}{2}\Delta_{\bm{\theta}}\Psi = \frac{1}{2}\left(\frac{\partial^{2}\Psi}{\partial\theta_{1}^{2}}+\frac{\partial^{2}\Psi}{\partial\theta_{2}^{2}}\right)
\label{eq:15}
\end{equation}
and the shear components $\gamma_{1}$ and $\gamma_{2}$ are given by
\begin{equation}
\gamma_{1} = \frac{1}{2}\left(\frac{\partial^{2}\Psi}{\partial\theta_{1}^{2}}-\frac{\partial^{2}\Psi}{\partial\theta_{2}^{2}}\right),\quad\gamma_{2} = \frac{\partial^{2}\Psi}{\partial\theta_{1}\partial\theta_{2}}, \quad\gamma=\sqrt{\gamma_{1}^{2}+\gamma_{2}^{2}}.
\label{eq:16}
\end{equation}
Points in the lens plane where
\begin{equation}
\det\mathcal{A}=(1-\kappa-\gamma)(1-\kappa+\gamma)=0,
\label{eq:17}
\end{equation}
form closed curves, the {\it critical curves}. Their corresponding image curves residing in the source plane are called {\it caustics}.
Images near critical curves can be significantly magnified and distorted, which, for instance, is indicated by the giant luminous arcs formed from source galaxies near caustics. Knowledge about the exact shape and location of these curves already allows one to make solid statements about the system's strong lensing properties. Since the possible critical curves and caustics of A2390 are relatively well constrained by observations of the here considered straight arc, estimating them for several density models in TeVeS will be one of the main tasks in this work.

\begin{figure}
\includegraphics[width=\linewidth]{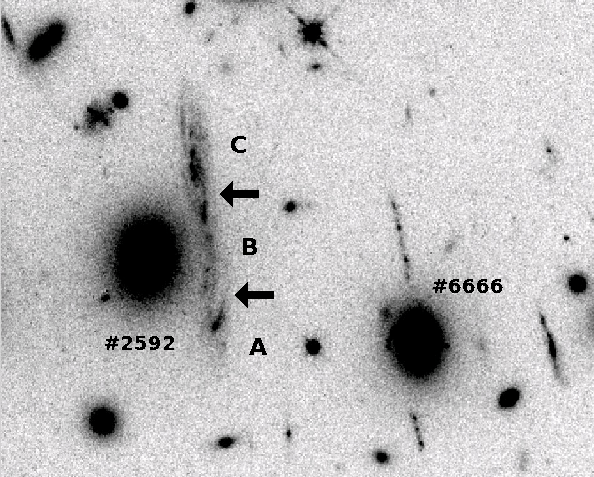}
\caption{A small section of an HST/WFPC2 observation of A2390 shows the impressive straight arc on the left side. Characterized by two breaks along its light profile (present in other observed bands as well), the arc can be decomposed into three segments labeled $A$, $B$ and $C$, respectively \cite{Pello1991}. Also visible are the galaxy $2592$, which is located adjacent to the arc, as well as the galaxy $6666$ (see Table \ref{table1}).
}
\label{figure0}
\end{figure}

\section{Observations of the galaxy cluster A2390}
\label{section3}
\subsection{X-ray gas and member galaxies of A2390}
\label{section31}

The galaxy cluster A2390 at redshift $z=0.23$ \cite{LeBorgne1991,Yee1996} is one of the richest and most luminous clusters known in the literature. Several interesting properties, e.g. the large abundance of lensing arcs and arclets \cite{Pello1991}, an elongated galaxy distribution \cite{Mellier1989} and its large velocity dispersion \cite{Carlberg1996}, have made the analysis of this system particularly attractive. In the context of GR, A2390 has been subject to extensive study by means of different techniques including virial (e.g., Ref. \cite{Benatov2006}), x-ray \cite{Bohringer1998,Allen2001,Hicks2006,Vikhlinin2006}, redshift-space caustic \cite{Diaferio2005} and both weak \cite{Squires1996,Dahle2006,Hoekstra2007,Bardeau2007,Umetsu2009} and strong \cite{a2390straight,Narashima1993,Pierre1996,Frye1998b} lensing studies.

Observations with CHANDRA exhibit a very concentrated and highly peaked x-ray emission, indicating a strong cooling flow which is centered on cluster's central cD galaxy \cite{Allen2001}. On large scales, the x-ray morphology has been found to be strongly elliptical with an overall position angle (PA) comparable to the main cluster direction in the optical (PA $=133^{\circ}$) \cite{Pello1991}. Here and below, the PA is defined as the angular offset of the major axis with respect to the north-south direction, being measured counterclockwise. The data provide evidence for an elongated x-ray morphology in the very central part, and suggest the existence of a substructure in the cluster gas located roughly $40''$ ($\sim 147$kpc) from the cluster center. The CHANDRA image further reveals large-scale cavities in the x-ray surface brightness extending approximately $400$kpc from the center, where a sharp break in the surface brightness profile is visible. As observed in several other clusters \cite{McNamara2005}, such cavities are likely produced by bubbles of radio plasma emitted by the central active galactic nucleus. Despite these irregularities and the appearance of a secondary gas peak, however, the x-ray observations indicate that the system as a whole is relatively regular and, to good approximation, dynamically relaxed. Thus, if one excludes the cluster's central part, the overall assumption of hydrostatic equilibrium appears as a reasonable one.

There are also several studies of individual galaxies within the cluster. For instance, the properties of the central cD galaxy have been examined using optical \cite{Smail1998}, infrared and radio observations \cite{Edge1999}. A large sample of $216$ confirmed cluster members based on photometric and spectroscopic information is presented in Ref. \cite{Abraham1996}. More recent observations include a selection of $48$ early-type member galaxies which has been used to investigate their evolutionary status \cite{Fritz2005}. We note that the available observational data will be important for building a realistic cluster model in TeVeS.

\subsection{The straight arc of A2390}
\label{section32}
Among several arcs and arclets, the cluster A2390 exhibits an unusual, strongly lensed straight arc (see Fig. \ref{figure0}) which is located approximately $38''$ ($\sim 140$kpc) from the central cD galaxy \cite{Pello1991}. This particular arc is unusual in the sense that, as it is both located in the outer core region and adjacent to a lens galaxy lying in between arc and cD galaxy, it would be expected to appear curved with respect to the massive cluster center or the closest galaxy. Along its light profile, the arc further exhibits two breaks in surface brightness, symmetrically located relative to the closest galaxy's center. Spectroscopic analysis of the arc revealed that it is actually the joint image of two different sources, one at redshift $z=0.913$ (corresponding to $B-C$ in Fig. \ref{figure0}) \cite{Pello1991} and the other at $z=1.033$ (corresponding to $A$) \cite{Frye1998a}. In addition, ISOCAM observations of the image segment $B-C$ indicate the presence of an active star forming region and support the scenario of two interacting source galaxies at $z=0.913$ \cite{Lemonon1998}. Nevertheless, the found straightness requires a rather special lens configuration (also see Sec. \ref{section521}).

Apart from the system A2390, there also exist other detections of (relatively) straight images which are typically well modeled from the visible distribution of bright galaxies helped by the central cluster potential \cite{Wambsganss1989,Mathez1992,Hattori1998}. As already pointed out in the literature \cite{a2390straight}, a similar approach for A2390 within the usual framework of CDM would require extremely high mass-to-light ratios for individual galaxies, and thus yields a rather unrealistic scenario. In recent years, several authors have considered possible lens models which aim at reproducing such a straight image, and a first attempt was performed in Ref. \cite{Pello1991}. For instance, the fold caustic of a single, highly elliptical cluster lens can be used to create a straight image \cite{Narashima1993}. Such a model gives a result comparable with the arc's morphology, but fails to explain infrared observations. Adopting a very large ellipticity of the central cluster profile, it was demonstrated how a cusp model may produce the desired elongated image morphology \cite{Frye1998b}; however, this solution seems incompatible with other lensing constraints of the system. Building on the existence of x-ray substructure in the arc's vicinity, the authors of Ref. \cite{Pierre1996} employed a two-component model using an elliptical cluster center with axis ratio $b/a=0.7$ to explain the arc. Despite a slight deviation at $\sim 1\sigma$ significance, the obtained x-ray temperature profile and the projected mass within $38''$ ($\sim 140$kpc) appear consistent with those derived from the observed x-ray luminosity \cite{Allen2001}.

It seems obvious that any suitable model needs substantial fine-tuning to form the necessary lens configuration for straight images. As a consequence, all of these models are extremely sensitive and unstable with respect to perturbations due to the closest galaxy or additional substructure in the intracluster medium (ICM). While this does not pose a problem {\it per se}, it is nevertheless interesting to look for models with improved stability. From a general analysis on how to form straight images \cite{a2390straight}, it has been concluded that the most likely configuration involves a dark mirror component of the nearest galaxy located on the opposite side of the arc, counterbalancing the effect of the visible galaxy. With the help of the central cluster profile, this yields a so-called {\it beak-to-beak model} which explains the observed straight arc and, if realized with such a ``dark galaxy,'' is sufficiently stable against local perturbations. Alternatively, there is also the possibility of a {\it lips catastrophe} \cite{a2390straight}, i.e. a lips caustic just emerged or just about to emerge in three-dimensional caustic space (for a demonstration of a lips catastrophe in A370 see Ref. \cite{Abdelsalam1998}). Since such a model requires the lensing convergence - equal to the projected matter density in GR only - to peak at the arc's position, however, it is not supported by observations.

\subsection{A challenge for TeVeS and hot dark matter}

Concerning the situation in TeVeS, we may already state that the ``dark galaxy'' approach, i.e. a nonluminous matter distribution of galactic size, cannot be achieved with our choice of $11$eV SN HDM. Assuming that these particles are relativistic and thermalized at the time of decoupling (just like for active neutrinos around a temperature of several MeV which is much larger than the considered mass) \footnote{Note that whether or not SNs decouple whilst in thermal equilibrium depends on the assumed model, production mechanism and parameters, e.g. the mixing to active neutrinos. Since the physical processes in the early universe are yet unknown, the relic distribution of SNs is quite uncertain. Here we choose a thermal distribution to obtain the desired cosmological properties as discussed in Refs. \cite{angussn,sncluster}}, their Fermi-Dirac distribution freezes in, and their phase-space density is constrained by the Tremaine-Gunn (TG) limit \cite{TG1979}. For instance, a HDM galaxy in TeVeS would have a typical phase-space occupation number (we neglect factors of $\pi$ and order unity)
\begin{equation}
\begin{split}
\frac{\hbar^{3}dN}{d^{3}x d^{3}p} &\sim \frac{M}{m_{\nu}}\left (\frac{\hbar}{m_{\nu}\sigma r_{C}} \right )^{3} \sim \frac{a_{0}^{2}\hbar^{3}}{Gm_{\nu}^{4}\sigma^{5}}\\
&\sim 10^{3}\left (\frac{m_{\nu}}{11{\rm eV}} \right )^{-4}  \left ( \frac{\sigma}{100{\rm km\ s}^{-1}} \right )^{-5},
\end{split}
\label{eq:estimate}
\end{equation}
which exceeds unity, and thus the TG limit for thermal relics, unless the HDM mass $m_\nu$ is much larger than $11$eV (e.g., $\sim 1$keV warm dark matter) and/or the structure's velocity dispersion $\sigma \gg 100$km s$^{-1}$, hence above the galactic scale. The estimate given in Eq. \eqref{eq:estimate} assumes that the structure's dense core is subject to the Newtonian regime ($\mu\sim 1$), which gives a core size $r_{C}\sim GM/\sigma^{2}$, and the total mass $M \sim \sigma^{4}/Ga_{0}$ can be well approximated within the ``deep-MOND'' limit ($\mu\sim\sqrt{y}$). Also note that moving to masses significantly larger than $m_{\nu}=11$eV would spoil the dynamics of MOND in galaxies and thus eliminate the use of such HDM in the first place \cite{sncluster}.

Therefore, a combination of HDM and modified gravity may, in principle, face a challenge in order to create observed effects of dark substructure. The TG phase-space bound not only applies to HDM substructure, but also to its global distribution within the cluster, which presents a well-posed and constraining general test of TeVeS or similar theories supplemented by an additional HDM component. As other realistic lens models for the straight arc \cite{Pierre1996} also suggest a substantial amount of dark substructure, a basic question is whether there are TeVeS lens models which are compatible with the TG bound for $11$eV SNs. Before we can address this point, however, we need a reliable way of modeling the straight arc in TeVeS. An approach into this direction will be discussed below.

In preparation for the following sections, we introduce the terminology and procedure used for two different kinds of lens configurations in our analysis:

\paragraph*{Quasiequilibrium configurations}
Here we consider configurations which are based on the assumption of hydrostatic equilibrium. Both the cluster gas and the (SN) HDM component are modeled by symmetric, central density distributions, the latter having a maximum phase-space density set by the TG limit which is inferred in a self-consistent way by considering the equation of state for a partially degenerate neutrino gas \cite{neutrinos2,sncluster}. In addition, we include substructure in the form of visible galaxies and further allow for perturbations of the central distribution (gas + HDM) which are modeled by the same density profile as the central one (corresponding to structure of equal scale). We then check whether such configurations can produce the observed straight image in TeVeS.

\paragraph*{Nonequilibrium configurations}
In this case, we allow for any HDM distribution which is capable of explaining the straight arc. This includes complex distributions with
multipeaked mass densities and concentrations of different scale. Although we outline a general approach to lens models in TeVeS, we restrict our analysis to a bimodal configuration based on a model in GR (cf. Table \ref{table2} below) whose components exhibit dispersions $\sigma\gtrsim 500{\rm km\ s}^{-1}$ and appear consistent with the crude estimate of Eq. \eqref{eq:estimate}, i.e. $\sigma \gtrsim 400 {\rm km\ s}^{-1} ({m_\nu /11{\rm eV}})^{4/5}$. Approximately treating each density peak as a symmetric equilibrium distribution of SNs, we investigate whether they satisfy the TG phase-space limit for $m_{\nu}=11$eV. For simplicity, we do not account for baryonic substructure (galaxies) in this context.

\begin{table*}
\caption{Positions, line-of-sight configurations and masses of individual galaxy components for the density model of A2390: At the cluster's redshift ($z=0.23$), an angular scale of $1''$ corresponds to approximately $3.7$kpc.}
\begin{ruledtabular}
\begin{tabular}{c c c c c c c c c}     
\noalign{\smallskip}
\multicolumn{5}{c}{}     & \multicolumn{2}{c}{Line-of-sight configuration} & \multicolumn{2}{c}{Projected stellar mass $M(<1.5'')$}
\tabularnewline
ID \footnote{Identifiers for galaxies are taken from Ref. \cite{Fritz2005}.} & $\theta_{x}$ & $\theta_{y}$ & $\xi_{x}$ & $\xi_{y}$ & $A$ & $B$ & $M_{1}$ & $M_{2}$
\tabularnewline
   & \multicolumn{2}{c}{($''$)} & \multicolumn{2}{c}{(kpc)} & \multicolumn{2}{c}{(kpc)} & \multicolumn{2}{c}{($10^{11}M_{\odot}$)}
\tabularnewline
\noalign{\smallskip}
\hline
\noalign{\smallskip}
{\#}$2180$ & $-48.21$ & $-16.98$ & $-178.04$ & $-62.71$ & $0$ & $+850$ & $2.02$ & $1.60$
\tabularnewline

{\#}$2592$ & $-34.29$ & $13.32$ & $-126.63$ & $49.19$ & $0$ & $+850$ & $3.51$ & $4.66$
\tabularnewline

{\#}$2619$ & $-13.04$ & $28.80$ & $-48.16$ & $106.36$ & $0$ & $+850$ & $1.09$ & $0.49$
\tabularnewline

{\#}$2626$ & $-34.62$ & $29.86$ & $-127.85$ & $110.27$ & $0$ & $-850$ & $1.40$ & $0.79$
\tabularnewline

{\#}$6666$ & $-50.25$ & $14.03$ & $-185.57$ & $51.81$ & $0$ & $-850$ & $2.89$ & $3.21$
\tabularnewline

Substructure \footnote{The given values roughly indicate the position of the x-ray substructure presented in Ref. \cite{Pierre1996}.} & $-37$ & $25$ & $-137$ & $92$ & - & - & - & -
\tabularnewline

Center & $0$ & $0$ & $0$ & $0$ & - & - & - & -
\tabularnewline
\noalign{\smallskip}
\end{tabular}
\end{ruledtabular}
\label{table1}
\end{table*}

\section{Quasiequilibrium model of A2390}
Because of the nonlinear relation of the TeVeS scalar field to the underlying matter distribution, we cannot work with projected quantities, but need to perform our calculations in three dimensions. This significantly complicates the lensing analysis of A2390 and requires knowledge about the cluster's three-dimensional matter density. A first approach to our problem is to consider cluster configurations which are based on the assumption of hydrostatic equilibrium.
\label{section4}
\subsection{Distribution of baryonic material}
\label{section41}
Using available data of x-ray gas \cite{Allen2001,Vikhlinin2006} and individual galaxies \cite{Fritz2005}, we have modeled the distribution of baryons in A2390. Here we shall briefly present the results which are relevant for the analysis in Sec. \ref{section52}. A detailed description of our procedure can be found in App. \ref{appendix1}.

\begin{figure}
\includegraphics[trim = 30 0 0 0,width=0.9\linewidth]{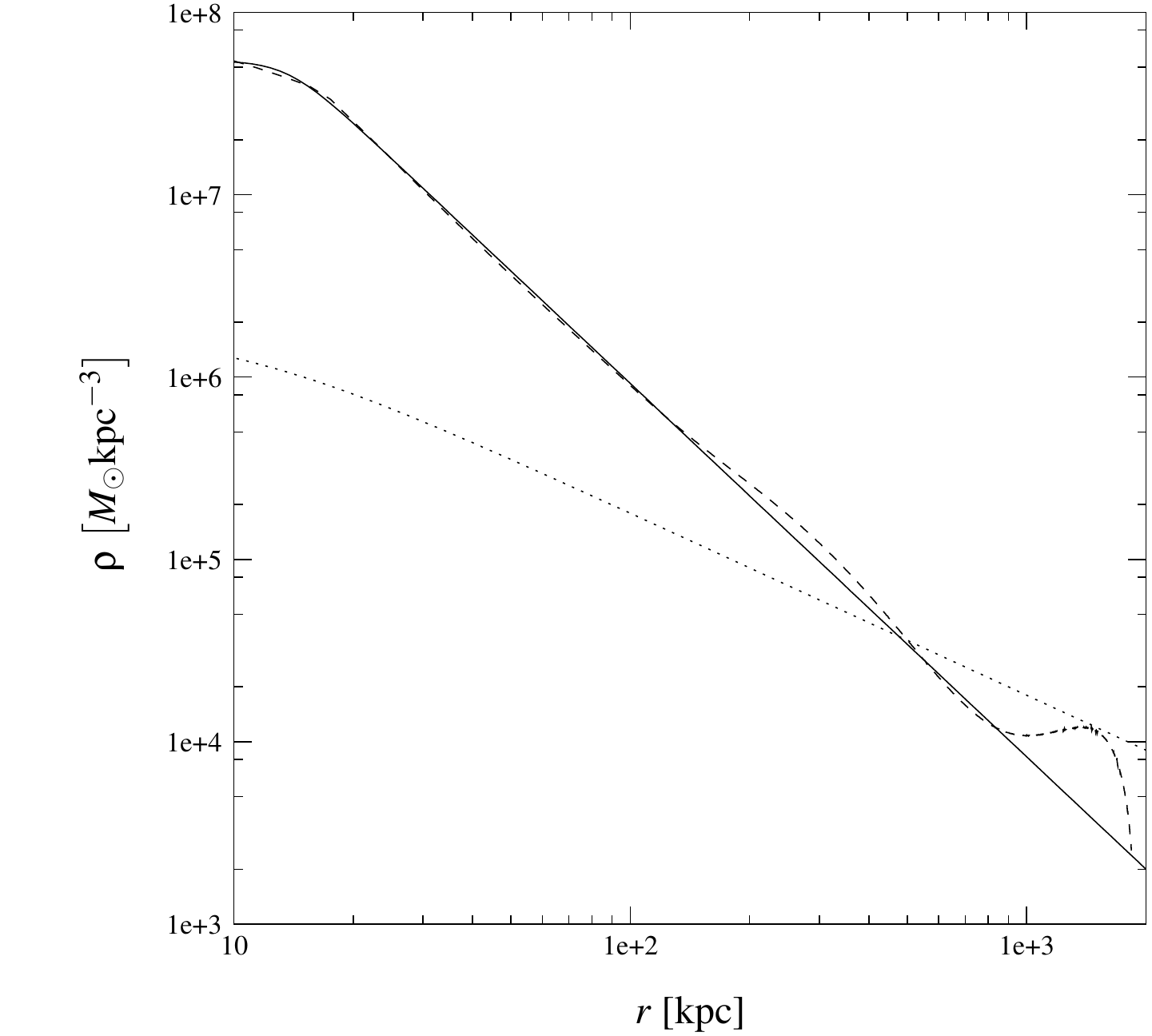}
\caption{TeVeS equilibrium configuration of $11$eV sterile neutrinos in A2390: The figure shows the calculated density distribution of neutrinos (dashed line), the analytic fit to this density using the profile specified in Eq. \eqref{eq:25} (solid line) and the central baryonic matter distribution derived from x-ray observations (dotted line).
}
\label{figure1}
\end{figure}

Figure \ref{figure1} shows the density distribution inferred from x-ray observations with CHANDRA (dotted line). In addition to this central profile, we consider the contribution of five massive early-type galaxies which are located close to the straight arc. The masses of these galaxies are derived following a twofold approach: The first estimate (denoted as $M_{1}$) is based on a direct conversion of observed luminosity to stellar mass while the second one ($M_{2}$) uses a dynamical method. In what follows, we shall consider both prescriptions and present results for the two different mass estimates below. We further assume that all galaxies can be described by a spherical density profile which is closely related to the Hernquist profile \cite{hernquist} for elliptical galaxies (see App. \ref{appendix12}). Using the notation of Ref. \cite{Fritz2005}, the basic properties of our models for the galactic components are illustrated in Table \ref{table1}.

\subsection{Adding massive neutrinos}
\label{section43}
As mentioned in Sec. \ref{section1}, TeVeS requires an additional matter component to consistently describe observations of galaxy clusters. Assuming $11$eV SNs within the original formulation of MOND, the authors of Ref. \cite{sncluster} derived their corresponding equilibrium density and (radial) velocity dispersion distributions for a sample of $30$ galaxy groups and clusters, including the system A2390. Starting from the observed density and temperature of the ICM \cite{Vikhlinin2006}, $\rho_{x}(r)$ and $k_{B}T_{x}(r)$, respectively, the assumption of hydrostatic equilibrium immediately allows one to determine the gravitational field as a function of radius:
\begin{equation}
g(r) = \frac{-k_{B}T_{x}(r)}{wm_{p}r}\left ( \frac{d\log\rho_{x}(r)}{d\log r} + \frac{d\log\ k_{B}T_{x}(r)}{d\log r}\right ),
\label{eq:20}
\end{equation}
where $w\approx 0.6$ is the mean molecular weight and $m_{p}$ the mass of the proton. The such derived result is typically accurate to $\sim 10$\% if equilibrium is realized \cite{Vikhlinin2006}. Using the above, one directly obtains the total enclosed MOND mass which is given by
\begin{equation}
M(r) = \frac{r^{2}g(r)\tilde{\mu}(x)}{G},\quad x=\frac{g}{a_{0}}.
\label{eq:21}
\end{equation}
Here $\tilde{\mu}$ corresponds to the MOND interpolating function defined in Eqs. \eqref{eq:1} and \eqref{eq:2}. Note that this is the only stage where the modification of gravity is involved. Once this function is specified, Eq. \eqref{eq:21} can be used to obtain the cluster's total density distribution, which then allows one to determine the contribution due to SNs by subtracting the known density of the ICM. 
Considering the equation of state for a partially degenerate neutrino gas, the resulting SN density $\rho_{\nu}$  is then used to infer the associated radial velocity dispersion $\sigma_{\nu}$ needed for equilibrium. A detailed description of the actual calculation can be found in Sec. $3$ of Ref. \cite{sncluster}. To check whether the results for $\rho_{\nu}$ and $\sigma_{\nu}$ are compatible with each other, one can exploit the TG phase-space constraint \cite{TG1979}. Assuming a Maxwellian velocity distribution, the maximally allowed density $\rho_{\nu,{\rm max}}$ for a given value of $\sigma_{\nu}$ reads
\begin{equation}
\rho_{\nu,{\rm max}} = \frac{g_{\nu}}{2}\frac{m_{\nu}^{4}}{(2\pi)^{3/2}\hbar^{3}}\sigma_{\nu}^{3},
\label{eq:22a}
\end{equation}
where the number of allowed helicity states is assumed as $g_{\nu}=2$ \cite{sncluster,Boyanovsky2008} and $m_{\nu}=11$eV.
For the ``simple'' MOND interpolating function which is defined as
\begin{equation}
\tilde{\mu}(x) = \frac{x}{1+x},
\label{eq:22}
\end{equation}
it has been found that the calculated SN phase-space density of all considered systems reaches the TG limit in the central part ($r\lesssim 20$kpc for A2390) \cite{sncluster}, meaning that the SNs acquire their densest possible configuration in that region. If the equilibrium assumption is valid, this result further implies that a small portion of the dynamical mass must be covered by the brightest cluster galaxy. As for the cD galaxy of A2390 and its contribution in this context, we refer the reader to App. \ref{appendix13}.

In principle, we could directly adopt the SN density of A2390 calculated in Ref. \cite{sncluster} for our simple cluster model if we specified a TeVeS free function $\mu$ which corresponds to the choice Eq. \eqref{eq:22} in MOND. For numerical reasons discussed in Ref. \cite{asymmetric} (hereafter Paper I) and to maximize possible MONDian effects, however, we assume a TeVeS free function of the following form:
\begin{equation}
\mu(y) = \frac{\sqrt{y}}{1+\sqrt{y}},
\label{eq:23}
\end{equation}
where $y$ is defined according to Eq. \eqref{eq:6a}. Apart from its simplicity, Eq. \eqref{eq:23} is close to Bekenstein's original choice of the free function (Paper I), and thus allows one to derive the TeVeS lens properties in a fully analytic way for certain configurations like, for example, spherically symmetric lens models \cite{lenstest}. In the intermediate and low acceleration regime, which is typically realized in galaxy clusters, the MONDian counterpart of Eq. \eqref{eq:23} can be expressed as \cite{teves}
\begin{equation}
\tilde{\mu}(x) = \frac{\sqrt{1+4x}-1}{\sqrt{1+4x}+1},
\label{eq:24}
\end{equation}
which is known to yield a less favorable description for the rotation curves of spiral galaxies than Eq. \eqref{eq:22} as it enhances gravity too efficiently \cite{freefunc}. Inserting the above into Eq. \eqref{eq:21}, we have repeated the analysis of Ref. \cite{sncluster} for A2390, and calculated the equilibrium SN density distribution suitable for our cluster model in TeVeS. The resulting density profile is shown as a dashed line in Fig. \ref{figure1}. Note that the apparent waviness is not a numerical artifact, but rather emerges from using the data of Ref. \cite{Vikhlinin2006} in Eq. \eqref{eq:20}. As the free function Eq. \eqref{eq:24} enhances gravity more efficiently than Eq. \eqref{eq:22}, the SN density is notably decreased (cf. Fig. $2$ of Ref. \cite{sncluster}), with the effect becoming stronger for larger radii. In the center, however, there is basically no change, indicating that the previous constraints due to the TG limit remain the same. To simplify the input into a numerical solver, the obtained SN density can be well fit by a profile of the following form:
\begin{equation}
\rho(r) = \frac{\rho_{0}}{\left (1+\left (r/r_{0} \right )^{\gamma} \right )^{1/4}},
\label{eq:25}
\end{equation}
where $\rho_{0} \sim 5.5\times 10^{7}M_{\odot}$kpc$^{-3}$, $r_{0} \sim 14$kpc and $\gamma \sim 8.2$. For comparison to the numerical result (dashed line), the analytic fit (solid line) is also illustrated in Fig. \ref{figure1}.

Note that the actual choice of the free function, which fixes the equilibrium distribution of SNs, will have no significant impact on the results for quasiequilibrium configurations presented in Sec. \ref{section52}. While the main cluster potential will almost be the same - it is exactly the same in case of spherical symmetry - for different $\mu$, only the effects of substructure, e.g. the contribution of individual galaxies in A2390, should be affected by the particular form of the function $\mu$. Therefore, our decision to use Eq. \eqref{eq:23} will result in optimistic estimates of effects intrinsic to the framework of TeVeS. Since we are interested in the regime of strong lensing, however, we expect these differences to be rather mild.

\begin{figure*}[t]
\begin{minipage}[t]{0.34\textwidth}
\begin{center} 
\includegraphics[trim = 20 0 0 0,width=\linewidth]{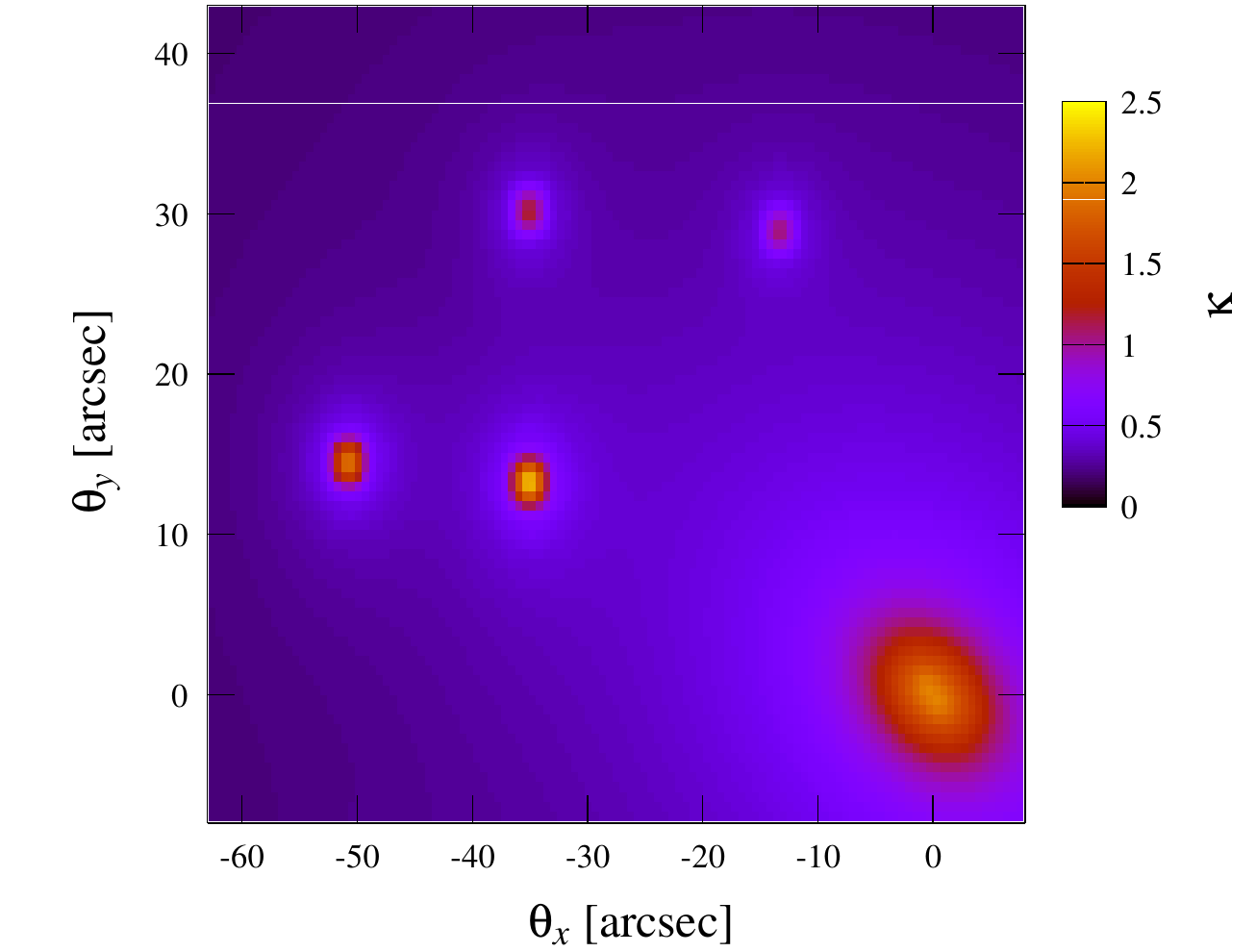}
\end{center}
  \end{minipage}
\hfill
\begin{minipage}[t]{0.34\textwidth}
\begin{center}
\includegraphics[trim = 20 0 0 0,width=\linewidth]{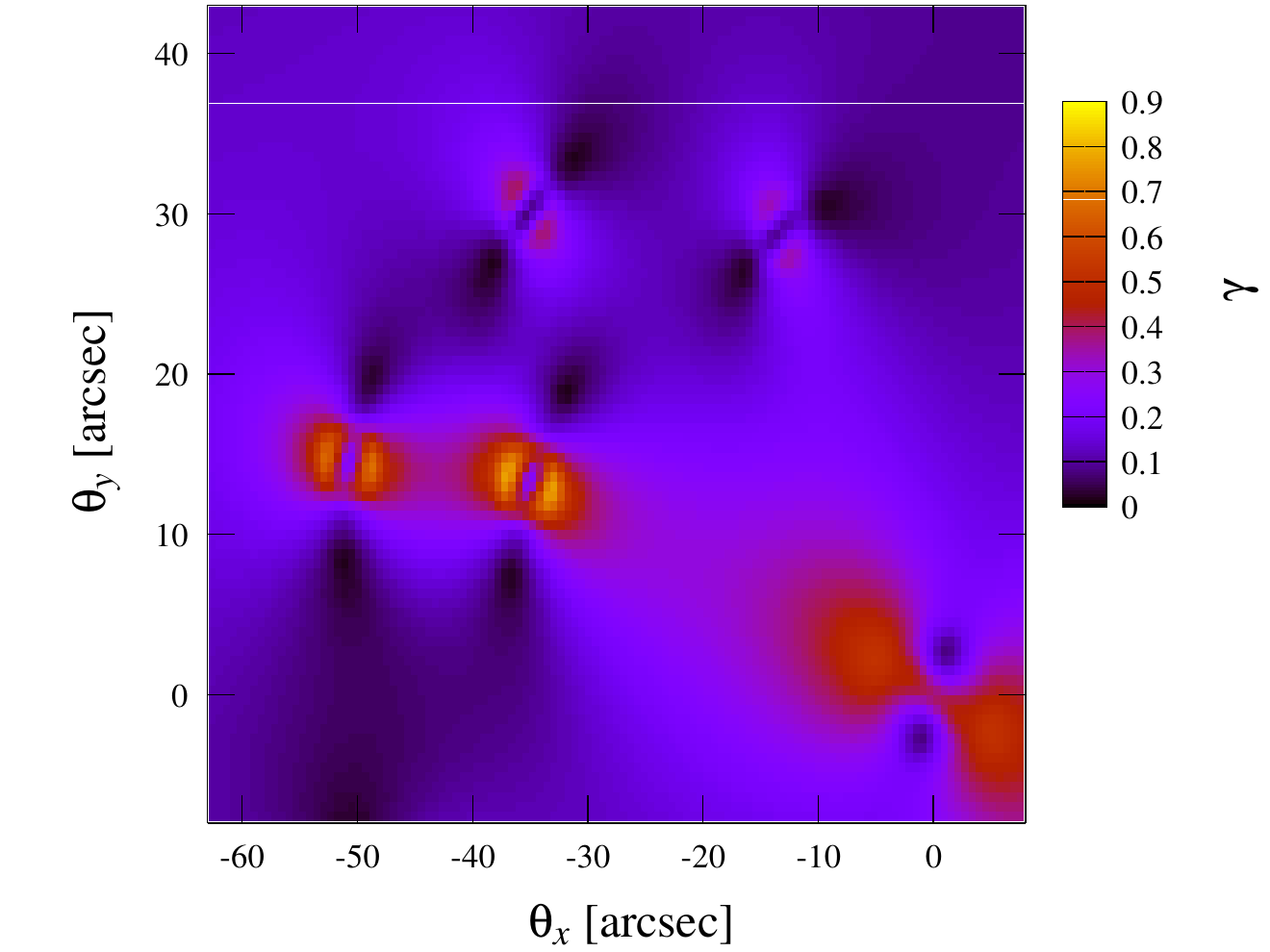}
\end{center}
\end{minipage}
\hfill
\begin{minipage}[t]{0.31\textwidth}
\begin{center}
\includegraphics[width=\linewidth]{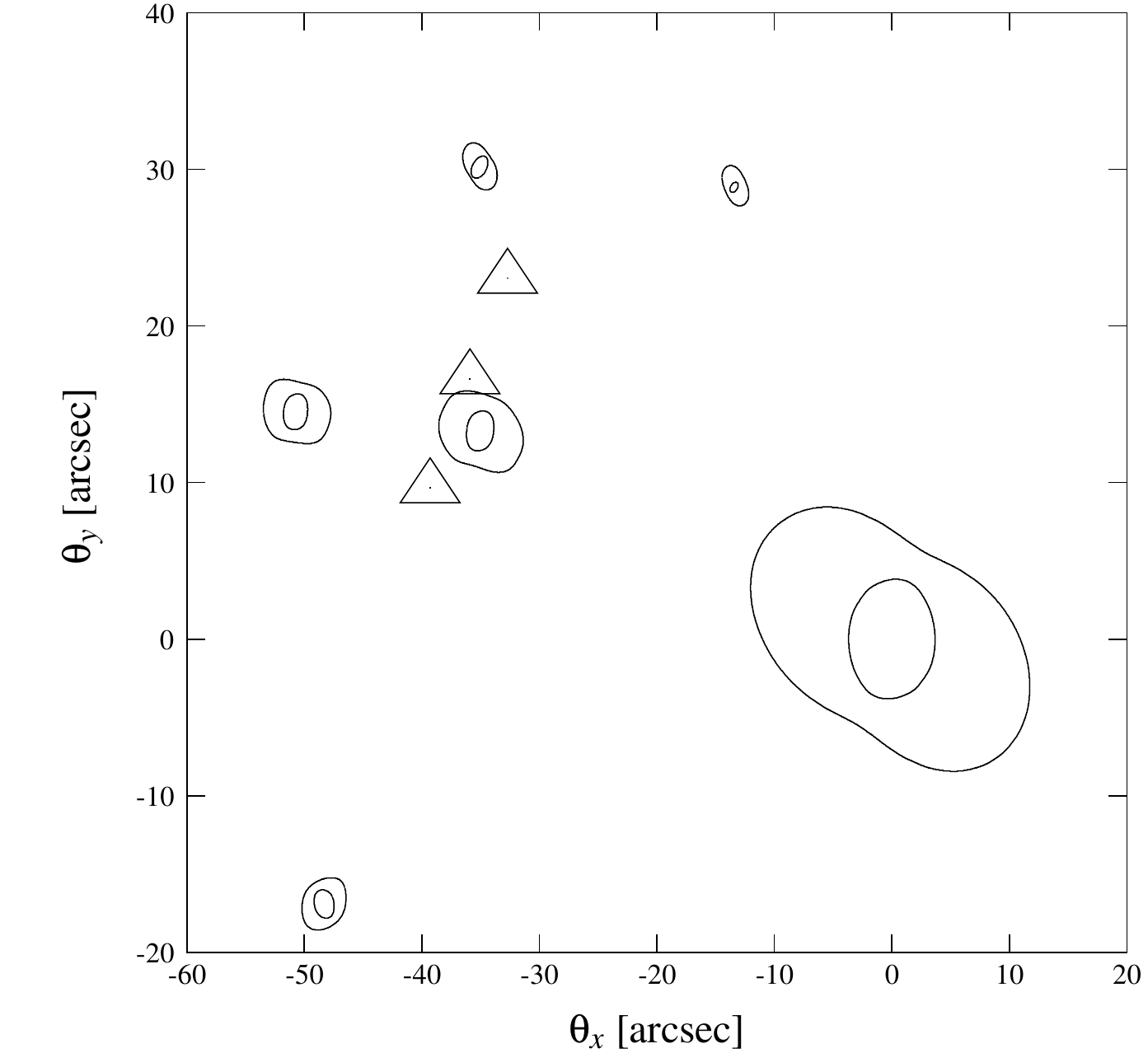}
\end{center}
\end{minipage}
\caption{TeVeS lensing maps for a quasiequilibrium configuration: Shown are the resulting convergence $\kappa$ (left), shear modulus $\gamma$ (middle), and critical curves for a cluster model with $e=0.7$, PA $=115^{\circ}$, mass model $M_{1}$ and line-of-sight configuration $B$. The triangles indicate the observed position of the straight arc.}
\label{figure4}
\end{figure*}

\begin{figure*}[t]
\begin{minipage}[t]{0.34\textwidth}
\begin{center} 
\includegraphics[trim = 20 0 0 0,width=\linewidth]{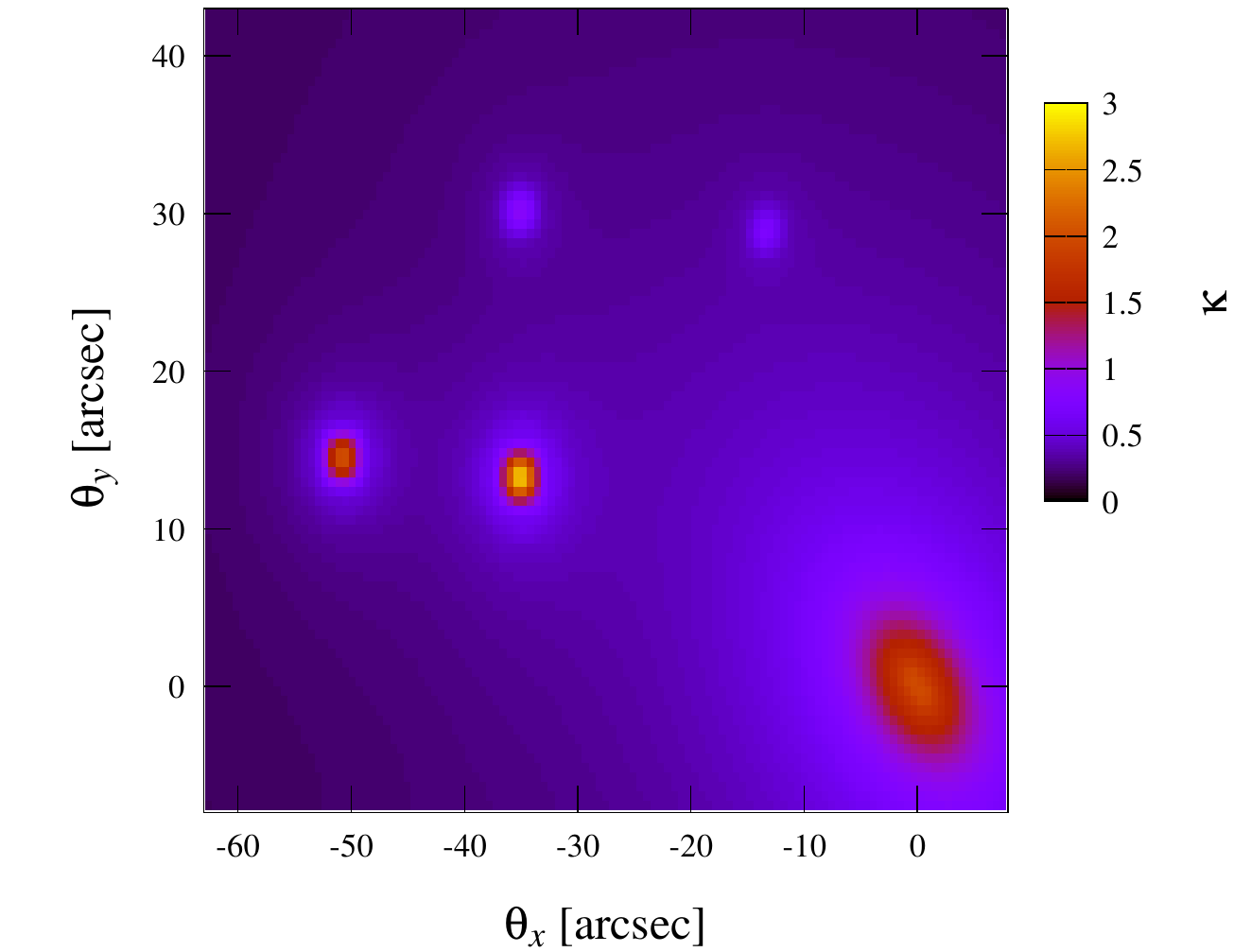}
\end{center}
  \end{minipage}
\hfill
\begin{minipage}[t]{0.34\textwidth}
\begin{center}
\includegraphics[trim = 20 0 0 0,width=\linewidth]{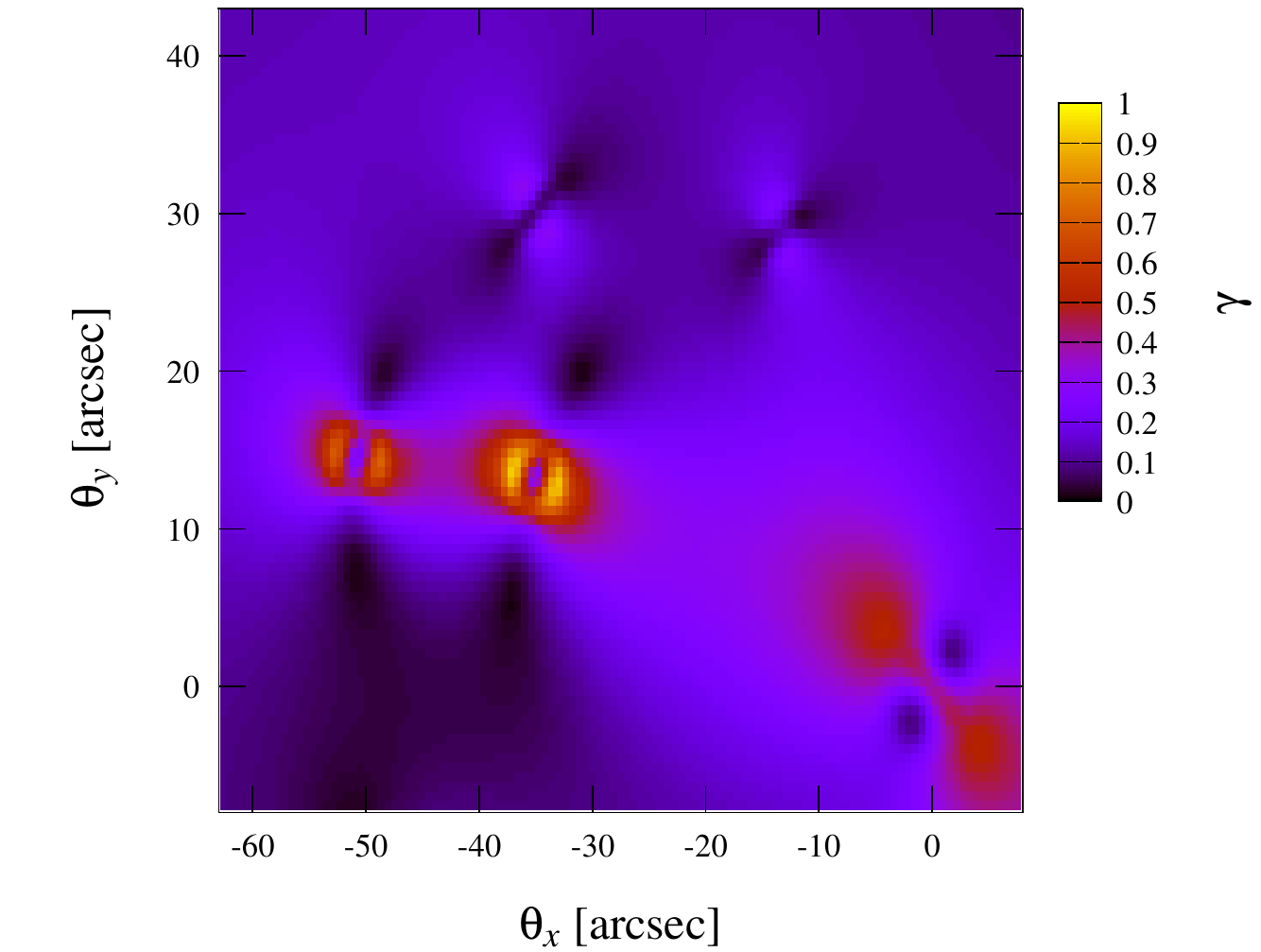}
\end{center}
\end{minipage}
\hfill
\begin{minipage}[t]{0.31\textwidth}
\begin{center}
\includegraphics[width=\linewidth]{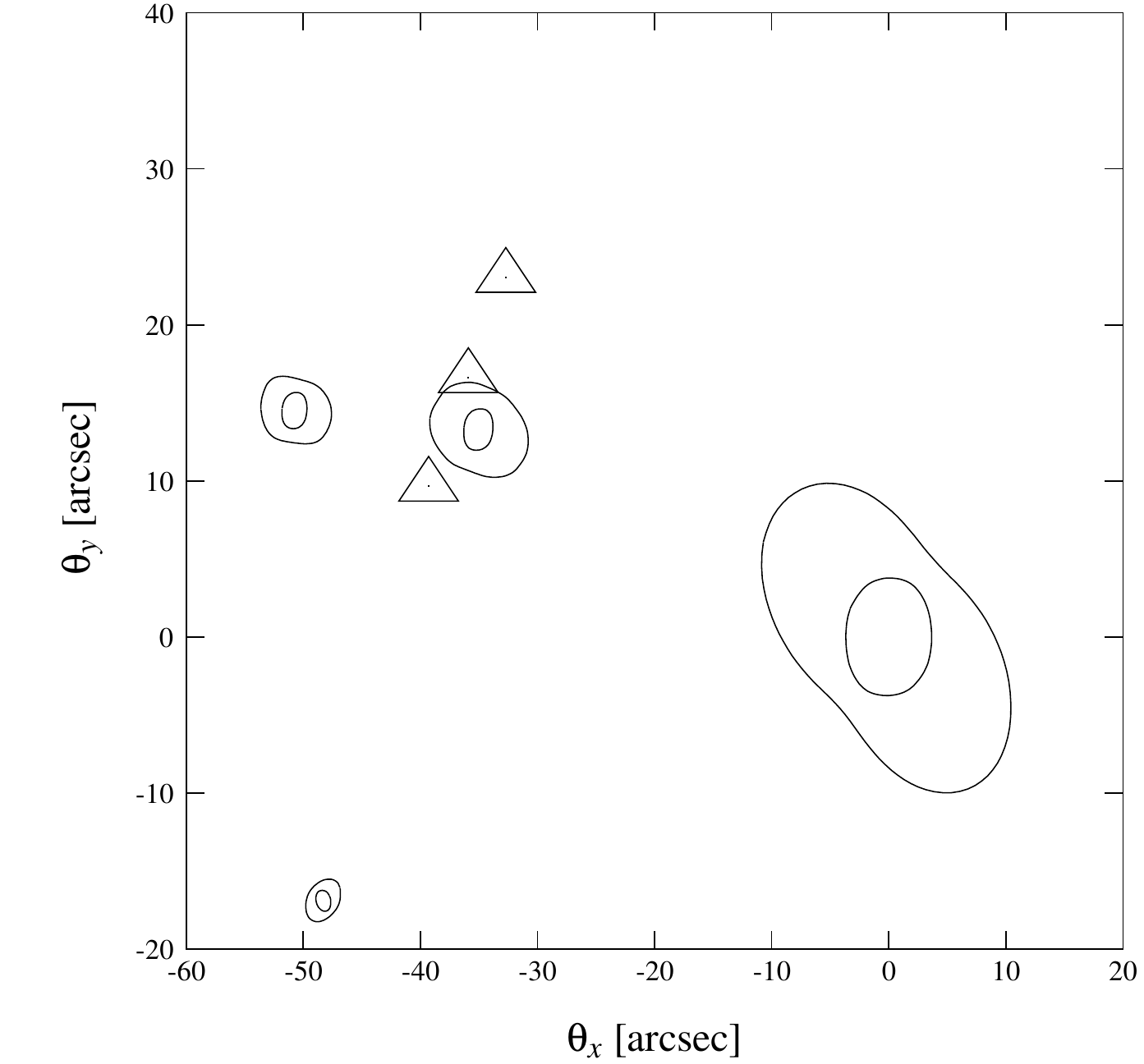}
\end{center}
\end{minipage}
\caption{Same as Fig. \ref{figure4}, but assuming PA $=133^{\circ}$ and mass model $M_{2}$.}
\label{figure5}
\end{figure*}

\section{Quasiequilibrium configurations}
\label{section52}
As a first approach, we shall investigate the strong lensing properties of quasiequilibrium configurations based upon several variations of the cluster model presented in Sec. \ref{section4}. Although these models do not provide a realistic description of the cluster's core region, their study will be extremely useful to explore intrinsic TeVeS effects and to see whether TeVeS offers alternative mechanisms - different from those in GR - which can produce straight images. For the sake of clarity, we discuss details on the used numerical tools and the basic simulation setup in App. \ref{appendix2}.
\subsection{Analysis of the TeVeS lensing maps}
\label{section521}
Considering the previously introduced equilibrium model of the cluster, we are still left with substantial freedom regarding the galaxies' line-of-sight positions which are not constrained by observations and may vary over the cluster's extent which we define by the model's cutoff radius $R=1$Mpc introduced in App. \ref{appendix11}. Also, to account for nonsphericity of the cluster, we shall allow an additional ellipticity for the central density distribution (x-ray and SNs) which is solely modeled within the observed plane. Together with a respective PA, this gives a total of $7$ free parameters for our simple model if we fix the galaxies' $M/L$ ratios. As for the range of ellipticities, we choose a maximum corresponding to an axis ratio of $b/a\sim 0.7$. Moving significantly beyond this threshold would cause a severe mismatch to x-ray observations \cite{Pierre1996,Allen2001}, thus yielding a rather unrealistic cluster description.

Modifications of the overall density profile along the line of sight have already been studied in Paper I: Varying the lens' extent between two extreme configurations, a disklike and a strongly ``cigar-shaped'' lens, can cause changes of up to $10-20$\% in the lensing maps as well as the critical curves. For realistic cluster models lying in a range between these extrema, however, this effect is expected to be less pronounced, typically accounting for deviations on the order of a few percent. Therefore, we shall ignore such modifications in this work. Also, since the straight arc's sources are located close in redshift space (the corresponding distances $D_{s}$ and $D=D_{d}D_{ds}/D_{s}$ only differ by roughly $3$\%), we restrict ourselves to a single source plane for our analysis. Unless otherwise stated, we will always work with a lens and source redshift of $z_{l}=0.23$ and $z_{s}=1$, respectively.

\begin{figure*}[t]
\begin{minipage}[t]{0.48\textwidth}
\begin{center} 
\includegraphics[trim = 40 0 0 0,width=0.8\linewidth]{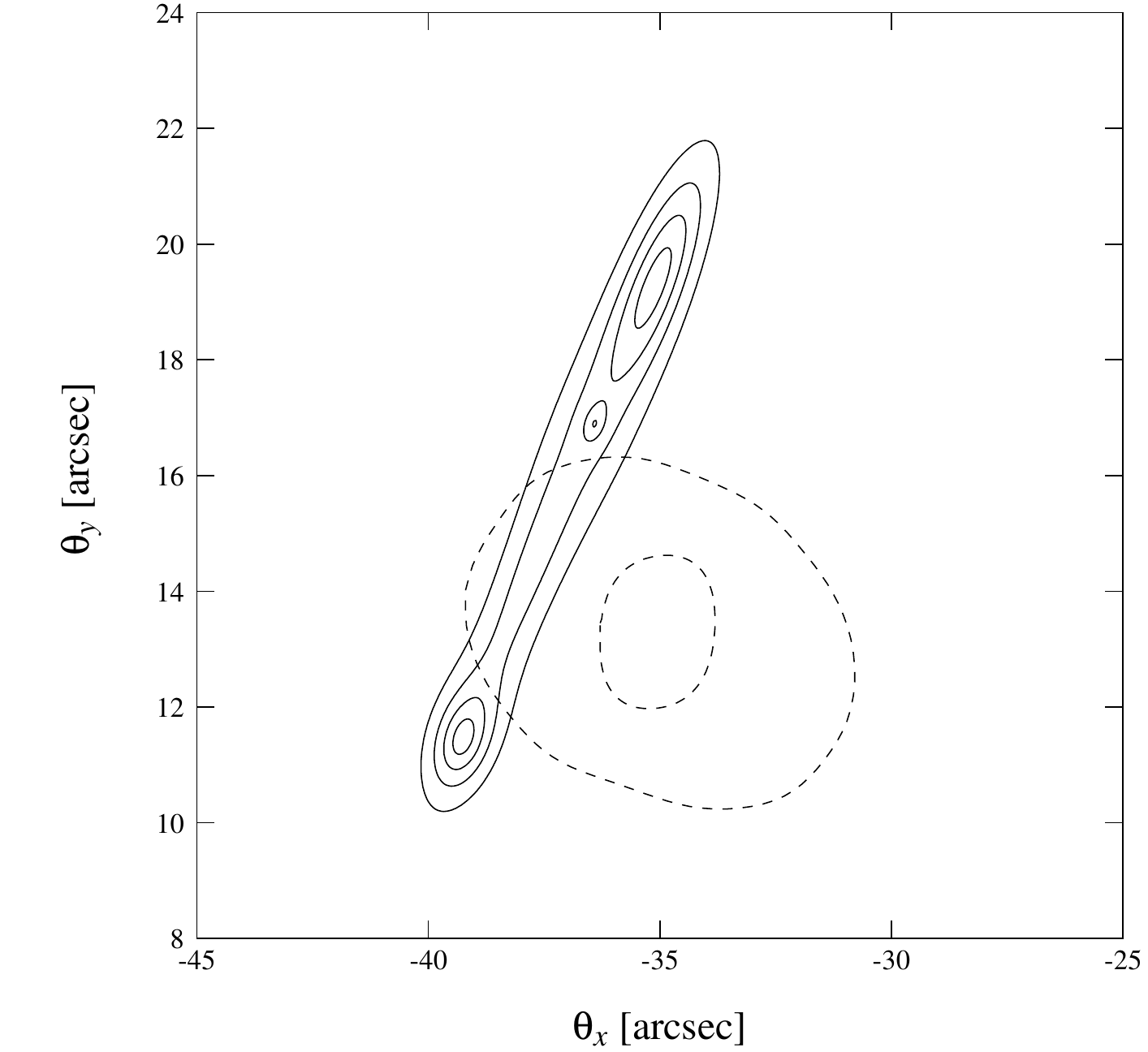}
\end{center}
  \end{minipage}
\hfill
\begin{minipage}[t]{0.48\textwidth}
\begin{center}
\includegraphics[trim = 40 0 0 0,width=0.8\linewidth]{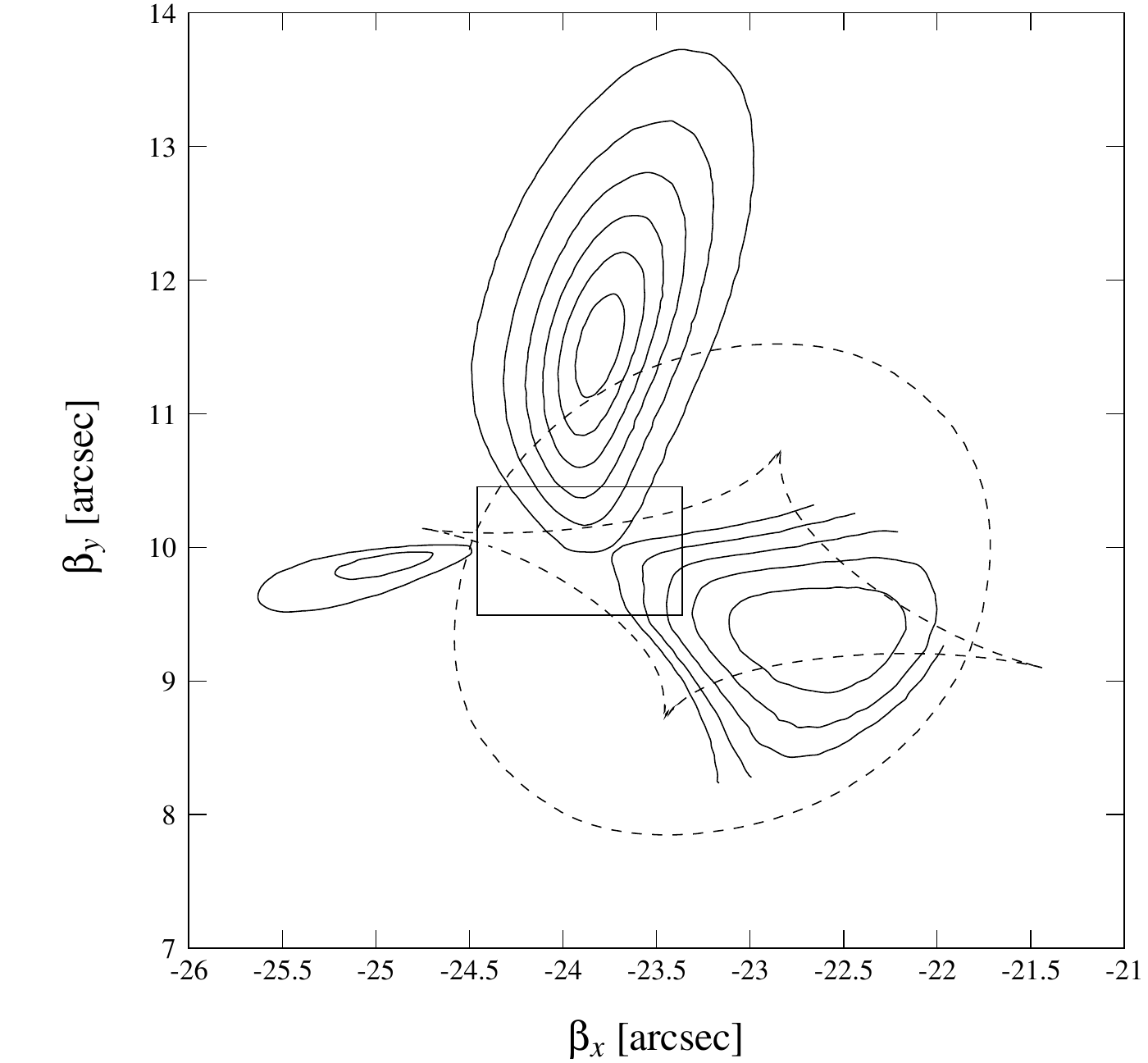}
\end{center}
\end{minipage}
\caption{Left: The generated image contours (solid lines), resembling the observed luminosity distribution of the straight arc, and the critical curves (dashed lines) for an equilibrium cluster model with $e=0.7$, PA $=133^{\circ}$, mass model $M_{2}$ and line-of-sight configuration $B$. Right: The resulting source distribution (solid lines) and lens caustics (dashed lines), where contours have been determined by averaging the calculated source points onto a regular grid. The open contour lines are due to a cutoff of the mapped image. In both panels, contours are in arbitrary units and chosen at equidistant levels.}
\label{figure5a}
\end{figure*}

For different plausible cluster configurations, we have found no quasiequilibrium model that is capable of producing (nearly) straight images at the observed arc's position. As all of our results are qualitatively very similar, we only present a selection of simulation runs in the following. For example, Fig. \ref{figure4} shows the calculated TeVeS lensing maps assuming an axis ratio $b/a\approx 0.7$ corresponding to an ellipticity of $e=0.7$, PA $=115^{\circ}$, mass model $M_{1}$ and line-of-sight configuration $B$ (see Table \ref{table1}). A similar case is illustrated in Fig. \ref{figure5}, assuming PA $=133^{\circ}$ and mass model $M_{2}$ while keeping all other parameters the same. We note that the lensing properties in the arc's vicinity are almost entirely dominated by the closest galaxy. The structure of the critical curves (right panel) already reveals that such models will produce strongly bent images with respect to the galaxy $2592$ at the position of interest. 
To elucidate this point and to demonstrate the problems of such configurations, we have constructed a luminosity distribution, which roughly resembles the observed image morphology. This distribution has then been mapped back into the source plane, assuming the ``second'' cluster model presented in Fig. \ref{figure5}. The generated image and its associated source distribution are shown in the left and right panels of Fig. \ref{figure5a}, respectively. Our particular example exhibits several features indicating that the model is not compatible with the observed straight image.
These features can be summarized as follows:
\begin{enumerate}[(a)]
\item Around the area where the three source patches visible in the right panel of Fig. \ref{figure5a} appear to intersect (marked with a rectangle), the inferred source distribution becomes multivalued. This remains true even after taking into account that the image is due to two distinct sources (see Sec. \ref{section32}), and thus the lens model turns out to be ambiguous and inconsistent.
\item Apart from the tangential caustic, i.e. the inner dashed line shown in the figure's right panel, the found source distribution also crosses the radial (outer) caustic, implying the existence of further images different from the straight arc. However, there is no evidence for such additional images as they are not observed in the system.
\item Assuming an average size of roughly $1''$ ($\sim 10$kpc) for galaxies at $z=1$, the source's constituents appear too big in angular size (up to $4''$), yielding a rather unlikely scenario. This problem further deteriorates if one tries to avoid the issues related to (a) and (b) by lowering the total mass of the nearby (lens) galaxy $2592$.
\end{enumerate}
Observations further indicate the presence of several faint elongated objects whose orientation is approximately the same as that of the arc, with a scatter of only a few degrees \cite{Pello1991}. Together with the above, these arclets strongly support the requirement for a special lens composition rather than the necessity for unusual source properties, suggesting that the lens configurations considered here are inappropriate to explain the straight image.

\begin{figure}
\includegraphics[width=\linewidth]{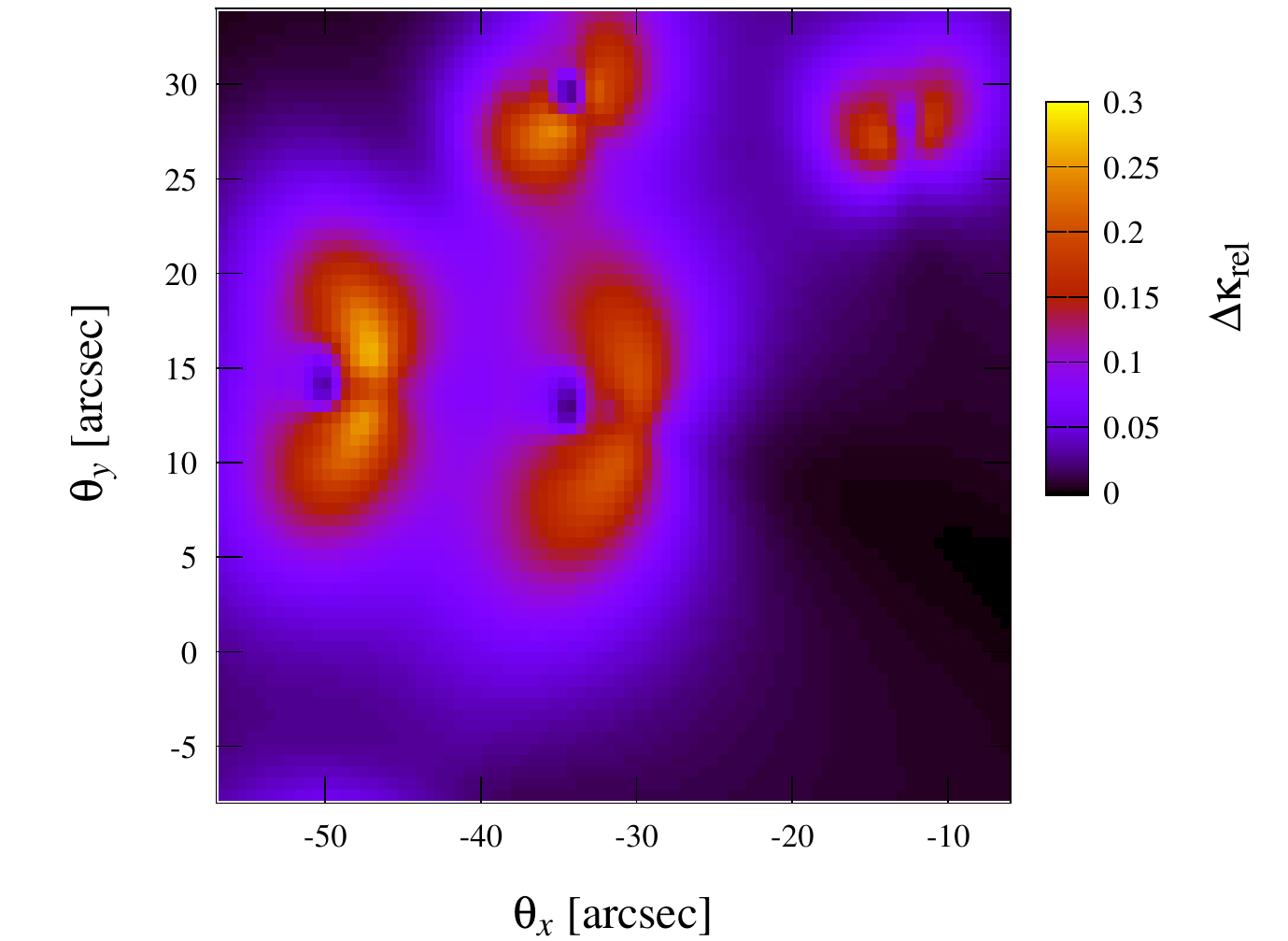}
\caption{Relative difference of the convergence maps calculated for the two line-of-sight configurations $A$ and $B$: The here presented result assumes $e=0.7$, PA $=133^{\circ}$, and mass model $M_{1}$.}
\label{figure6}
\end{figure}

This is the basic result of all simulated cluster models, which seems to be insensitive to the used mass model ($M_{1}$ or $M_{2}$) or the actual line-of-sight alignment of galactic components. To quantify the effect of the latter, we compared the lensing maps of individual models for two extreme line-of-sight configurations, $A$ and $B$. Adopting the parameters of the realization presented in Fig. \ref{figure4}, Fig. \ref{figure6} displays the obtained relative difference between the corresponding convergence maps. As we can see, the deviation can reach values up to $\sim 30$\% in regions of low (effective) surface density, but remains smaller ($\lesssim 15$\%) in regions where $\kappa\gtrsim 1$. A comparison of the corresponding critical curves and caustics of galaxies reveals that this line-of-sight effect typically affects their position on the order of $\sim 10$\%, which has no qualitative impact on our results. As for the dependence on the actually used mass models (galaxies), we will investigate the influence of varying $M/L$ ratios in the next section.

\subsection{Variation of mass-to-light ratios}
\label{section522}
So far, we have restricted our analysis to two sets of $M/L$ ratios for the cluster's galaxies (see Table \ref{table1}). How robust are our results with respect to variations of these ratios? Here we take a simplified approach to obtain reasonable estimates of the effect on the strong lensing properties, in particular, the critical curves. In what follows, we use a Hernquist profile \cite{hernquist} with fixed core radius $r_{H}=3$kpc for the density distribution of galaxies [corresponding to the limit $\epsilon =0$ in Eq. \eqref{eq:19}]. In the isolated case and for our choice of the free function $\mu$, this allows one to express the lensing properties fully analytically (e.g., Paper I), and the deflection angle is given by
\begin{equation}
\hat\alpha(\xi) = \frac{r_{H}A(\xi)}{\sqrt{|\xi^{2}-r_{H}^{2}|}}\left(\frac{4\xi\sqrt{GMa_{0}}}{r_{H}}+\frac{4GM\xi}{\xi^{2}-r_{H}^{2}}\right)-\frac{4GM\xi}{\xi^{2}-r_{H}^{2}},
\label{eq:30}
\end{equation}
where
\begin{equation}
A(\xi) = 
\begin{cases}
\arcsinh{\sqrt{\left|1-\left(r_{H}/\xi\right)^{2}\right|}} & \xi<r_{H}\\
\arcsin{\sqrt{1-\left(r_{H}/\xi\right)^{2}}} & \xi>r_{H}
\end{cases}.
\label{eq:31}
\end{equation}
Furthermore, we will assume that
\begin{enumerate}[(a)]
\item the superposition principle remains valid; i.e. the lensing maps of isolated galaxies and the cluster background can just be added, which is rigorously true if the components are infinitely separated from each other and leads to an optimistic estimate otherwise, and that
\item the cluster background at each galaxy's position can be modeled as an external contribution with locally constant convergence $\kappa_{C}$ and shear modulus $\gamma_{C}$.
\end{enumerate}
Choosing polar coordinates, the latter yields an effective cluster deflection potential of the following form:
\begin{equation}
\Psi(\theta,\varphi) = \frac{\kappa_{C}}{2}\theta^{2}+\frac{\gamma_{C}}{2}\theta^{2}\cos(2(\varphi - \varphi_{0})),
\label{eq:32}
\end{equation}
where the external shear's principle axes system is defined by $\varphi_{0}$. Locally, the system's total shear modulus, relevant for the determination of critical curves and caustics, depends nonlinearly on the contribution due to the Hernquist lens and the cluster,
\begin{equation}
\gamma^{2}_{\rm tot} = (\gamma_{1,H} + \gamma_{1,C})^{2} + (\gamma_{2,H} + \gamma_{2,C})^{2}.
\label{eq:33}
\end{equation}
Because of the shear's tensor property, the above is anisotropic, which directly affects the resulting position of critical curves given by Eq. \eqref{eq:17},
\begin{equation}
(1 - \kappa)^{2}-\gamma^{2} = 0.
\label{eq:34}
\end{equation}
To obtain the mean effect due to the cluster background, we perform an average over $\varphi$ and all possible orientations of the external shear field, which leads to
\begin{equation}
\overline{\gamma^{2}} = \gamma_{H}^{2} + \gamma_{C}^{2}.
\label{eq:35}
\end{equation}
\begin{figure}
\includegraphics[trim = 70 0 -40 0,width=\linewidth]{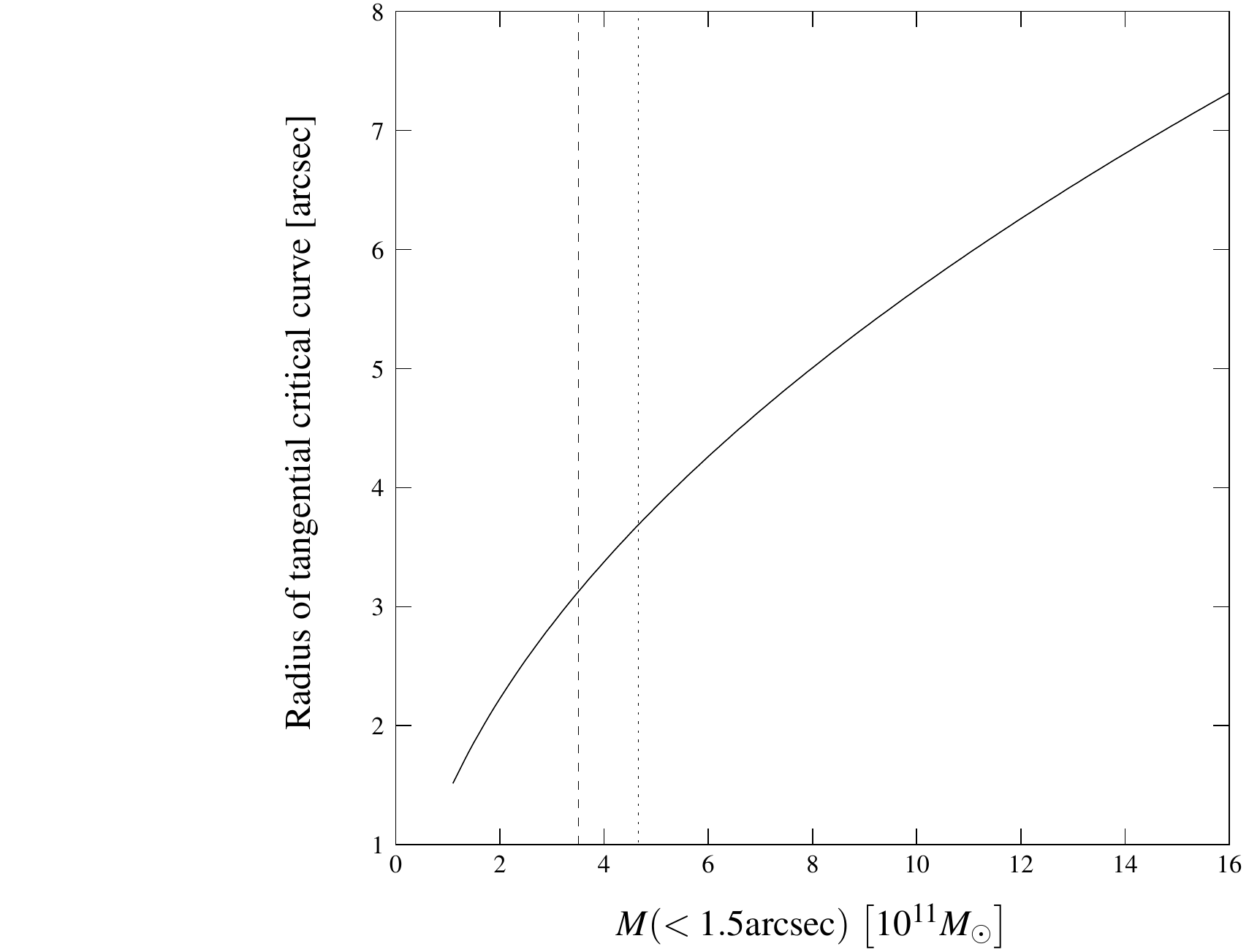}
\caption{Predicted mean radius of tangential critical curves of a Hernquist lens embedded into the external cluster field: The mean radius is plotted as a function of the projected enclosed mass within an aperture of $3''$ diameter ($\sim 11$kpc). The vertical lines indicate the values of the galaxy $2592$ for mass model $M_{1}$ (dashed line) and $M_{2}$ (dotted line), respectively.}
\label{figure7}
\end{figure}
We use the simulation result for an equilibrium cluster model with $e=0.7$, PA $=115^{\circ}$ and no galaxies to estimate the parameters of the background model. Around the arc's position, this roughly fixes $\kappa_{C}\approx 0.29$ and $\gamma_{C}\approx 0.17$. For this case, Fig. \ref{figure7} shows the resulting mean radius of the tangential critical curve as a function of the enclosed galactic mass within an aperture of $3''$ diameter. While this should give a reasonable picture for the galaxy $2592$, which resides close to the arc, the such estimated radii will be too large for the other galaxies. These are located in regions where the background has a weaker impact ($\kappa_{C}$ and $\gamma_{C}$ take lesser values), leading to an optimistic prediction of their mean critical-curve size. Assuming small variations in $M/L$, the figure suggests no qualitative changes of our previous results. Even if we consider that $M/L$ ratios may change up to a factor of $4$ in the infrared, we find a maximum increase of the mean critical-curve radius corresponding to a factor of approximately $2$. At most, such an extreme scenario could come close to a merged-cusp model for the galaxies $2592$ and $6666$, but this configuration cannot explain the arc due to its inappropriate orientation and position of critical curves and caustics in the lens and source plane, respectively. We therefore infer that the basic result of Sec. \ref{section521} does not depend on the particularly assumed $M/L$ ratios of individual galaxies - unless the background potential is substantially modified.

\begin{figure*}[t]
\begin{minipage}[t]{0.48\textwidth}
\begin{center} 
\includegraphics[trim = 20 0 0 0,width=\linewidth]{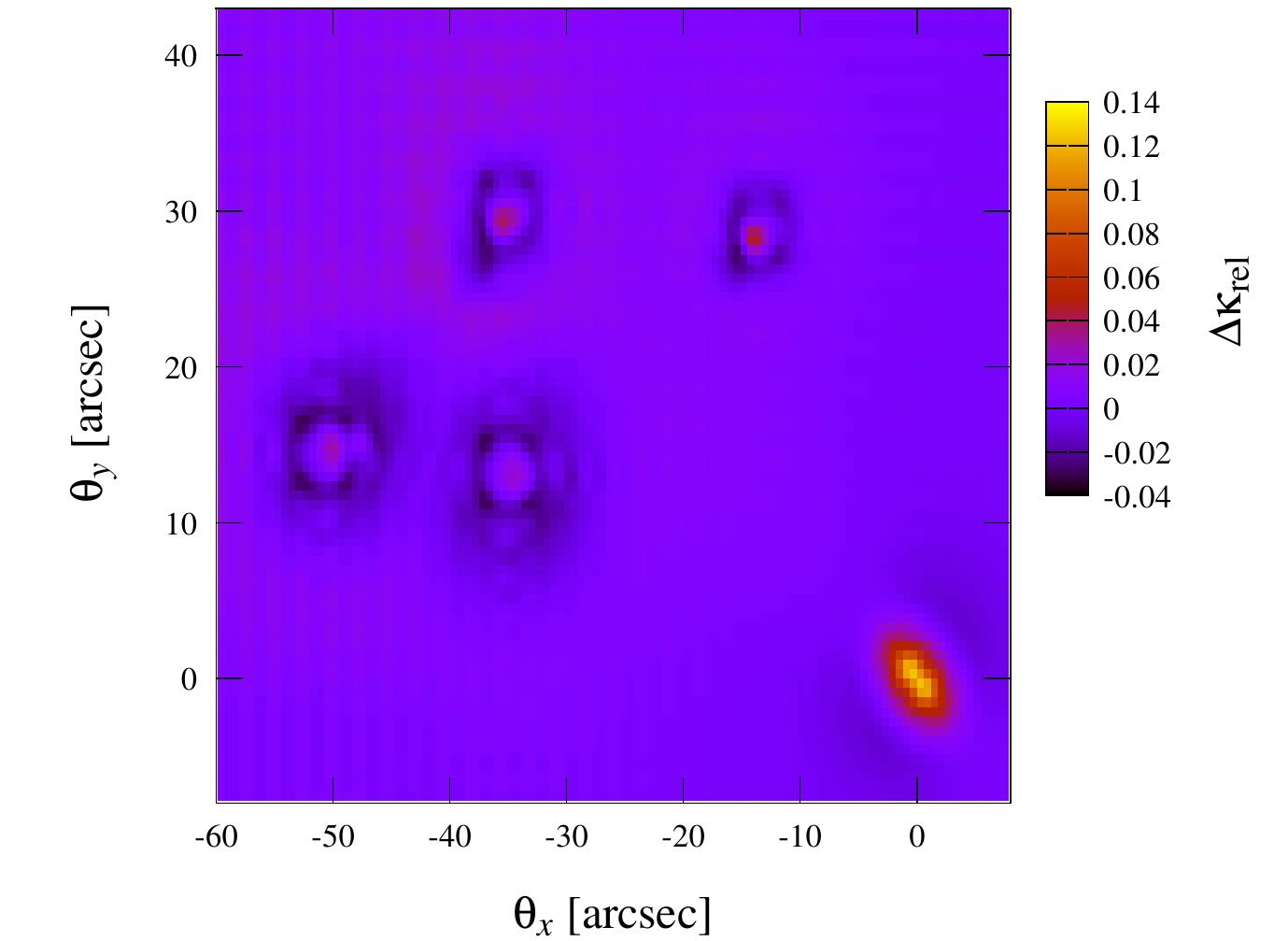}
\end{center}
  \end{minipage}
\hfill
\begin{minipage}[t]{0.48\textwidth}
\begin{center}
\includegraphics[trim = 20 0 0 0,width=\linewidth]{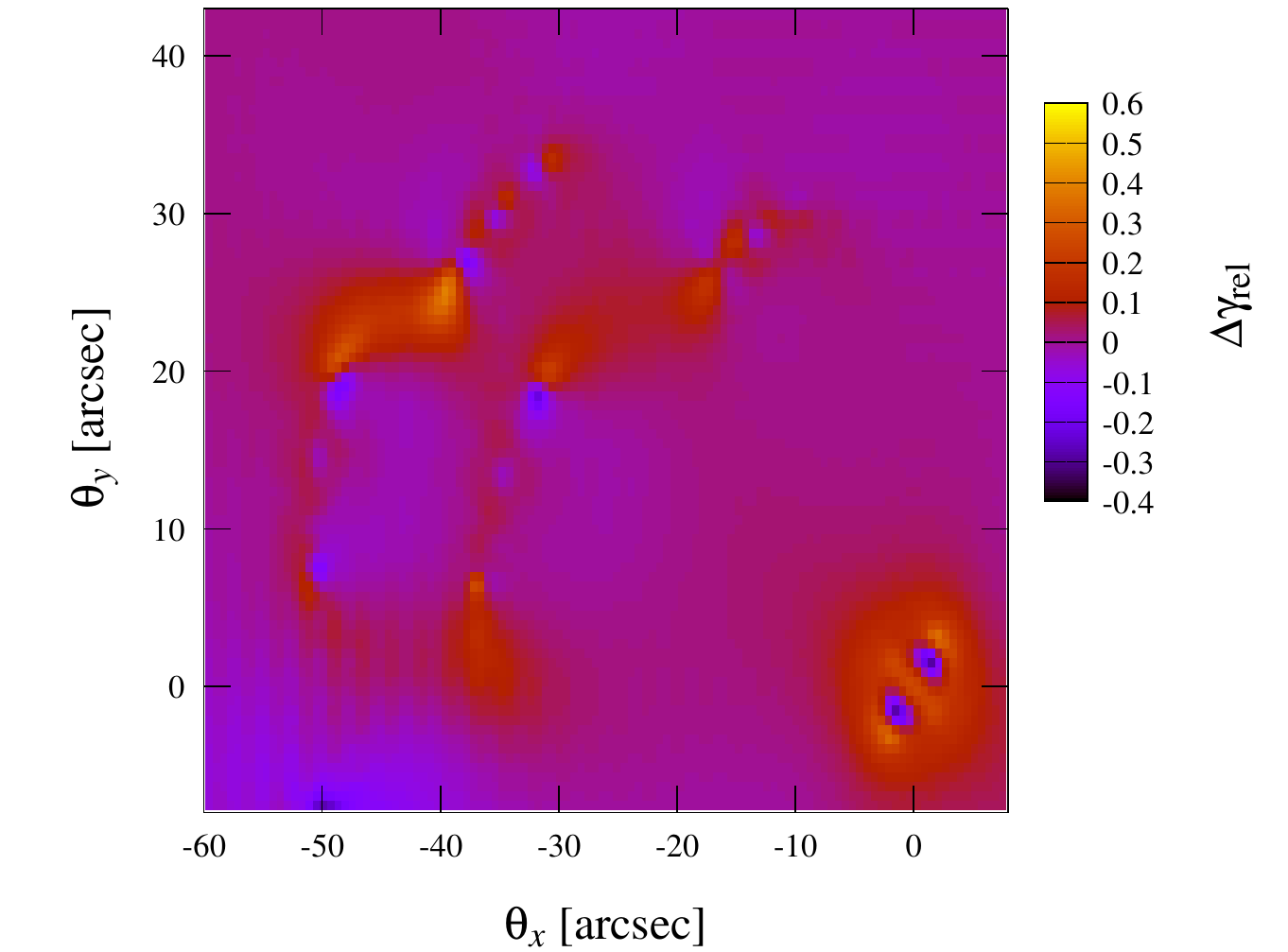}
\end{center}
\end{minipage}
\caption{Simulation results with proper treatment of the scalar field equation compared to those where the curl field $\mathbf{h}$ has been set to zero: The figure illustrates the relative deviation in the corresponding lensing maps, the convergence $\kappa$, and the shear modulus $\gamma$, assuming an equilibrium model with $e=0.7$, PA $=133^{\circ}$, mass model $M_{1}$, and line-of-sight configuration $B$. The visible gridlike structure is a combined effect of Fourier fluctuations, interpolation, and the division by values close to zero.}
\label{figure8}
\end{figure*}

We have also explored the influence of perturbations to the central cluster profile. For this reason, we have assumed a secondary spherical clump made out of gas and SNs which follows the same profile as the central distribution and accounts for $10-15$\% of the system's total mass. The clump's position has been chosen from a narrow range roughly centered on the detected substructure in the x-ray map \cite{Pierre1996,Allen2001} (see Table \ref{table1}). Again, we have found no qualitative difference compared to previous simulations. The calculated deviations in the lensing maps are on the order of a few percent, leaving a basically negligible impact on the critical curves and caustics. Similar statements apply to an overall increase of the central density profile by $10-20$\%.

Together with the results presented in Sec. \ref{section521}, we thus conclude that TeVeS quasiequilibrium configurations with $11$eV SNs are not capable of explaining the observed arc. In particular, we find no evidence for the formation of beak-to-beak or lips catastrophes \cite{a2390straight} due to intrinsic TeVeS effects, which could give rise to straight images. Therefore - just as in GR - a suitable TeVeS lens model needs substantially more mass as well as a special density distribution in the cluster's core region. A general procedure on how to obtain such models will be discussed in the next section.

\section{Nonequilibrium configurations}
\label{section53}
In the following, we shall outline a general approach for modeling cluster lenses in TeVeS which allows one to use existing GR lens models to estimate the needed TeVeS lens properties. Adopting a bimodal lens model for the straight arc, we will present an example of such a lens and discuss implications for the modified framework.
\subsection{Systematic approach to cluster lenses}
\label{section531}
Taking a naive point of view, one might expect that strong lensing is subject to the strong acceleration regime, and therefore it should be enough to consider the limit $\mu\rightarrow 1$. In this case, all relevant equations would reduce to their GR counterparts, allowing a conventional lensing analysis. Previous calculations (Paper I) have shown that such an approximation is not justified. In particular, the scalar field can have a significant impact on the second derivatives of the lensing potential. For instance, this can increase the radii of critical curves by up to a factor of $2$, depending on the assumed mass distribution of the lens (cf. Figure $7$ of Paper I). As we shall see below, however, there is another way of simplifying the lensing problem in TeVeS.

\begin{figure*}[t]
\begin{minipage}[t]{0.48\textwidth}
\begin{center} 
\includegraphics[trim = 40 0 0 0,width=0.8\linewidth]{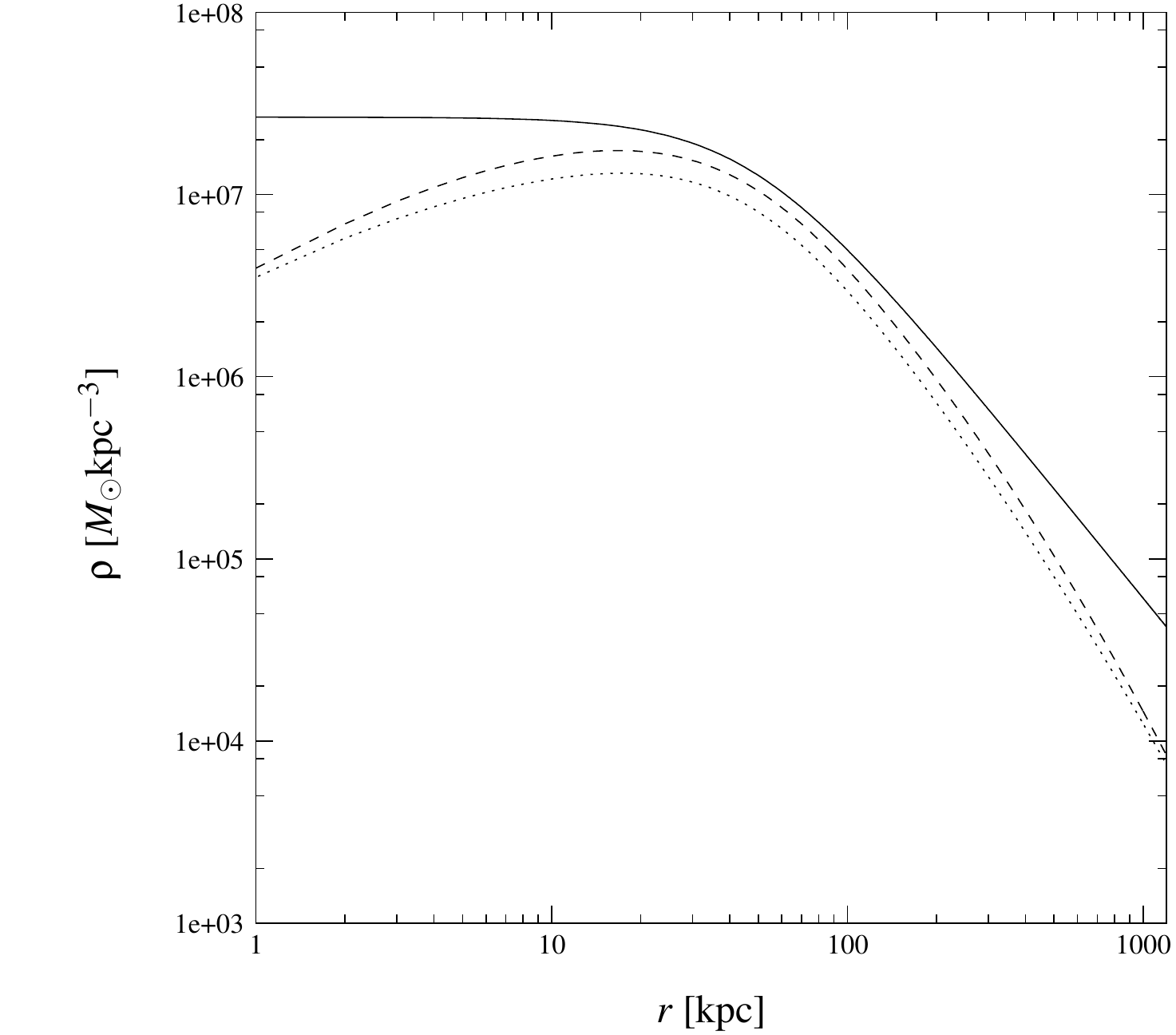}
\end{center}
  \end{minipage}
\hfill
\begin{minipage}[t]{0.48\textwidth}
\begin{center}
\includegraphics[trim = 40 0 0 0,width=0.8\linewidth]{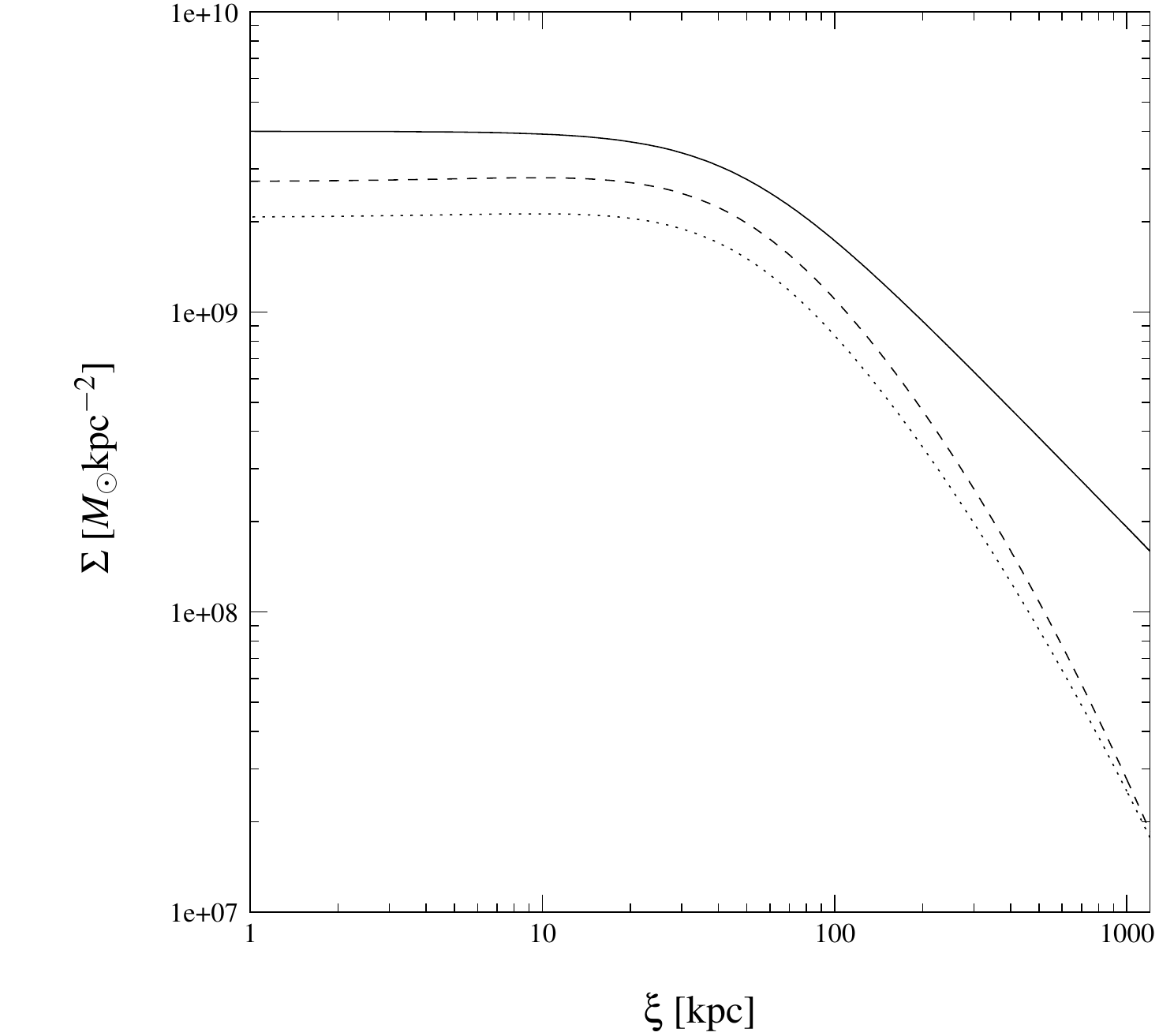}
\end{center}
\end{minipage}
\caption{Spherically averaged density (left panel) and corresponding projected surface density (right panel) profiles for the bimodal lens model in TeVeS: Shown are the results for the Newtonian dynamical mass (solid line) and the two TeVeS free interpolating functions corresponding to Eq. \eqref{eq:41} (dotted line) and Eq. \eqref{eq:42} (dashed line), respectively. Note that the three-dimensional mass density profiles are entirely dominated by the contribution of SNs within a radius of a few hundred kpc. For $r\lesssim 25-30$kpc, the derived densities are well below the TG-limit-saturating equilibrium distribution of $11$eV SNs in Fig. \ref{figure1}, and they are also much broader. This already indicates that the main component's phase-space limit is not violated here.}
\label{figure9}
\end{figure*}

Let us return to the scalar field given by Eq. \eqref{eq:5}. Integrating once, we may recast this equation as
\begin{equation}
\mu\bm{\nabla}\phi = \frac{k}{4\pi}\left (\bm{\nabla}\Phi_{N}+\bm{\nabla}\times\mathbf{h} \right ),
\label{eq:36}
\end{equation}
where $\mathbf{h}$ is a regular vector field determined up to a gradient by the condition that the curl of the right-hand side of Eq. \eqref{eq:36} must vanish. We note that the main difficulty associated with solving the scalar equation are the generally nonvanishing components of $\mathbf{h}$. If for any reason $\mathbf{h}\approx\mathbf{0}$, Eq. \eqref{eq:36} reduces to a relatively simple algebraic relation between the gradients of the scalar and the Newtonian potential,
\begin{equation}
\mu\bm{\nabla}\phi \approx \frac{k}{4\pi}\bm{\nabla}\Phi_{N},
\label{eq:37}
\end{equation}
which can easily be inverted by numerical means to give $\bm{\nabla}\phi$, assuming that the Newtonian potential (or only its gradient) are known. Therefore, we want to address the question of how the field $\mathbf{h}$ is affecting the corresponding lensing maps in the strong lensing regime. We already expect $\mathbf{h}$ to be important around local extrema of the Newtonian potential, but it is difficult to make any intuitive guesses about its quantitative impact in stronger gravity regions as well as on the final projected result. The most straightforward approach to this problem is a direct comparison of simulations treating the full scalar equation to those where $\mathbf{h} = \mathbf{0}$. To this end, we have taken our previous quasiequilibrium models and fed them into a modified version of our solver, now assuming Eq. \eqref{eq:37} to determine the scalar gradient. Since our choice of $\mu$ is very close to that presented in Ref. \cite{lenstest} (e.g., Paper I), our code assumes
\begin{equation}
\left |\bm{\nabla}\phi\right | = \sqrt{a_{0}\left |\bm{\nabla}\Phi_{N}\right |}
\label{eq:38}
\end{equation}
to calculate $\bm{\nabla}\phi$. For instance, adopting an equilibrium model with $e=0.7$, PA $=133^{\circ}$, mass model $M_{1}$ and line-of-sight configuration $B$, the relative deviation of the lensing maps from the proper solution is presented in Fig. \ref{figure8}. While the convergence varies by $5-15$\%, differences in the shear map can be as high as $\sim 50$\%. As expected, the largest deviations occur in regions where the Newtonian gradient approaches the null vector, which, for example, can be seen in the very core of the central elliptic profile for both the convergence and shear maps.

Clearly, the impact of the curl field $\mathbf{h}$ is not negligible in regions of low gravity. Concerning the domain of strong lensing, however, we find the following: Comparing the corresponding critical curves and caustics, the curl field turns out to be much less important. Interestingly, the obtained deviation with respect to their position in the lens and source plane, respectively, is only about $\lesssim 2-3$\%. Within a sufficiently large environment around these curves accounting for all strong lensing features, the accuracy of the approximated ($\mathbf{h} = \mathbf{0}$) lensing maps is typically of the same order, meaning that the curl field negligibly contributes to the strong lensing properties of a given matter distribution. Our result appears to generally hold for strong cluster lenses and indicates that it is enough to consider Eq. \eqref{eq:37} in the context of TeVeS lens models. Therefore, if one specifies the line-of-sight extent of the total system as well as individual matter components, this offers a direct systematic way of modeling strong lenses in TeVeS.

\subsection{Modeling the straight arc in TeVeS}
\label{section532}
Based on the result of Sec. \ref{section531}, one could, in principle, take an available GR lens fitting routine, modify it to include the TeVeS scalar field according to Eq. \eqref{eq:37}, and use it to obtain a lens model for the straight arc. It is obvious that such an approach will be computationally more demanding because the scalar's contribution has to be evaluated in three dimensions, and one also needs to invoke numerical integration to derive the desired projected quantities. In the following, however, let us consider an alternative way to estimate the necessary deflection mass and its distribution in TeVeS. For this reason, we start from the bimodal GR model derived in Ref. \cite{Pierre1996} which, in addition to the central matter clump, assumes a smaller subcomponent at approximately $45''$ ($\sim 166$kpc) from the cluster center. The second clump is motivated by the existence of substructure in the cluster's x-ray map which is used to infer its position in the lens plane. Both clumps are chosen to follow a pseudoisothermal elliptic mass distribution (PIEMD) \cite{Kassiola1993}, but the subcomponent's profile is assumed to be spherically symmetric. Correcting for the here used cosmological background, the model gives an enclosed projected mass of $M_{a} \sim 1.2\times 10^{14}M_{\odot}$ within a circular aperture of $38''$ ($\sim 140$kpc) radius from the cluster center. As typical for strong lensing mass models, this estimate should lie within $\sim 30$\% of the true value \cite{Schneider2006}.

Using the arguments presented in Sec. \ref{section531}, it is obvious that there exists an analogous bimodal lens model in TeVeS. To obtain a spherically averaged density estimate in TeVeS, we ignore the secondary clump, which negligibly contributes to the enclosed mass within the given aperture, and also assume that the main component can be described by a spherically symmetric density profile. Thus, its three-dimensional matter distribution can be written as
\begin{equation}
\rho(r) = \rho_{0}\frac{r_{C}^{2}}{r_{C}^{2}+r^2},
\label{eq:39}
\end{equation}
where $\rho_{0}$ is the central density and $r_{C}$ the core radius. Alternatively, Eq. \eqref{eq:39} may be written in terms of its asymptotic velocity dispersion $\sigma_{\infty}$ associated with the density profile of a singular isothermal sphere:
\begin{equation}
\rho(r) = \frac{1}{2\pi G}\frac{\sigma_{\infty}^{2}}{r_{C}^{2}+r^2}.
\label{eq:39a}
\end{equation}
The corresponding enclosed mass of this density distribution at radius $r$ reads
\begin{equation}
M(r) = 4\pi r_{C}^{2}\rho_{0}\left (r-r_C \arctan\left (\frac{r}{r_C}\right ) \right ).
\label{eq:40}
\end{equation}
Since our choice of the free function allows us to make use of Eq. \eqref{eq:38}, it is possible to express the enclosed mass in TeVeS, which effectively generates the same dynamical mass as Eq. \eqref{eq:39}, as
\begin{table}
\caption{Fiducial parameters of the bimodal lens configuration presented in Ref. \cite{Pierre1996}: Here the subclump is offset by approximately $45''$ ($\sim 166$kpc) from the main component.}
\begin{ruledtabular}
\begin{tabular}{c c c c c}     
\noalign{\smallskip}
& $b/a$ & PA & $r_{C}$ & $\sigma_{\infty}$
\tabularnewline
   &  & ($^{\circ}$) & ($''$) & (km s$^{-1}$)
\tabularnewline
\noalign{\smallskip}
\hline
\noalign{\smallskip}

Central main clump & $0.71$ & $49.2$ & $12\pm 5$ & $950\pm 100$
\tabularnewline

Subclump & $1$ & $-$ & $7-12$ & $420-500$
\tabularnewline
\noalign{\smallskip}
\end{tabular}
\end{ruledtabular}
\label{table2}
\end{table}
\begin{equation}
\begin{split}
M_{\rm eff}(r) &= M(r) + \frac{a_{0}r^2}{2G}\left (1-\sqrt{1+\frac{4GM(r)}{a_{0}r^2}}\right )\\
&= M(r)\frac{4s}{\left (1+\sqrt{1+4s}\right )^{2}},
\end{split}
\label{eq:41}
\end{equation}
where $s=GM(r)/a_{0}r^{2}$. As previously noted, however, the choice of Eq. \eqref{eq:23} does not yield a good description of galaxy rotation curves. Adopting a TeVeS free function corresponding to Eq. \eqref{eq:22}, a similar calculation leads to
\begin{equation}
M_{\rm eff}(r) = \frac{\left\lbrack M(r)\right\rbrack^{2}}{M(r)+ a_{0}r^2/G} = M(r)\frac{s}{1+s}.
\label{eq:42}
\end{equation}
Setting $r_C \approx 13''$ ($48$kpc) \cite{Pierre1996} and requiring that the enclosed projected dynamical mass within $38''$ is still given by $M_{a}$, the above expressions can be used to derive the underlying density distributions which, together with the resulting surface density profiles, are illustrated in Fig. \ref{figure9}. The visible density drop-off within $r\lesssim 20$kpc is a consequence of the assumed PIEMD and probably unphysical, but can easily be avoided by changing the central profile in favor of a peaked and finite core, fixing the enclosed mass around $r=140$kpc (and thus keeping the lens properties needed for the arc). Of course, our results depend on the assumed line-of-sight extent specified by Eq. \eqref{eq:39}, but the derived surface densities should vary by only a few percent for different models (see Sec. \ref{section521}). We also note that the ``modified'' density profiles yield a finite mass; for both Eqs. \eqref{eq:41} and \eqref{eq:42}, the total mass is given by (taking the limit $s\rightarrow 0$)
\begin{equation}
\lim\limits_{r\to\infty}M_{\rm eff}(r) = \frac{16\pi^{2}Gr_{C}^{4}\rho_{0}^{2}}{a_{0}}.
\label{eq:43}
\end{equation}
Although the profiles are not diverging, the relevant mass typically extends to large radii. Therefore, the use of such profiles within the full TeVeS solver is very inefficient because very large box sizes would be necessary to perform the calculations, which underlines the advantage of neglecting the curl field for strong lensing models (see Sec. \ref{section531}).

\begin{figure}
\includegraphics[trim = 40 0 0 0,width=0.85\linewidth]{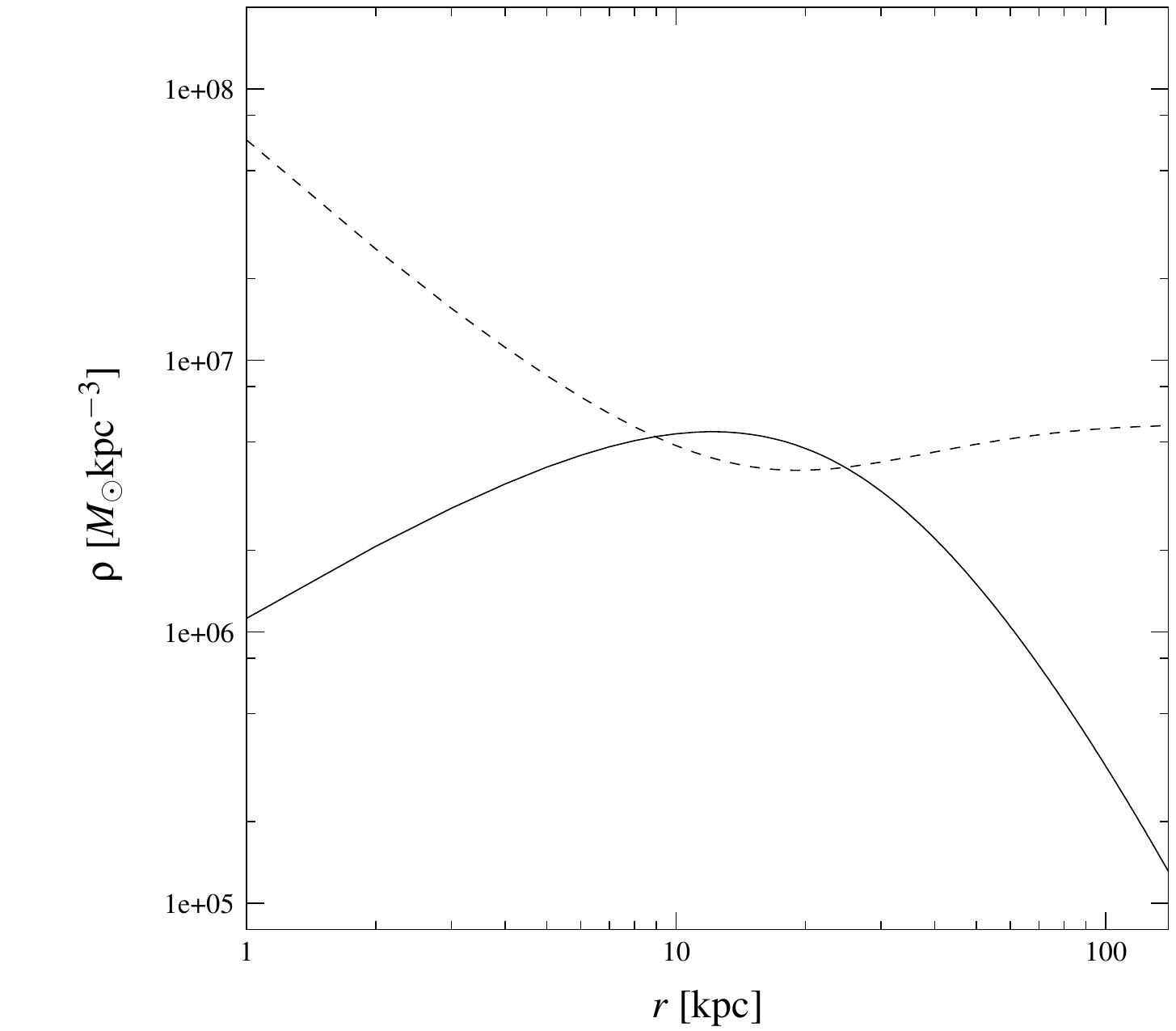}
\caption{Estimated $11$eV SN density distribution (solid line) and corresponding TG limit (dashed line) for a subclump model with $r_{C}\approx 7''$ ($26$kpc) and $\sigma_{\infty}\approx 500$km s$^{-1}$: The origin is centered on the subcomponent, and the TG limit has been calculated according to Eq. \eqref{eq:22a}, following the prescription of Ref. \cite{sncluster}. Note that the slight ``wiggly'' feature of the dashed line is due to a nonuniform dispersion $\sigma(r)$ which is computed in a self-consistent way \cite{sncluster}.}
\label{figure10}
\end{figure}

The resulting total lensing mass is entirely dominated by SNs within a radius of a few hundred kpc, which allows one to ignore the contribution of gas and stellar material to excellent approximation. We have checked that the three-dimensional density distributions in Fig. \ref{figure9}, basically representing the lens model's main component, are consistent with the TG limit estimated for hydrostatic equilibrium and a Maxwellian velocity distribution, following the approach of Ref. \cite{sncluster}. This is already indicated by the fact that the derived densities are much broader and well below the TG-limit-saturating $11$eV SN equilibrium distribution (shown in Fig. \ref{figure1}) for $r\lesssim 25-30$kpc. At the arc's position ($\theta =38''$), the actual enclosed projected mass of the TeVeS lens models is given by $6.1\times 10^{13}M_{\odot}$ or $8.0\times 10^{13}M_{\odot}$, assuming Eq. \eqref{eq:41} or \eqref{eq:42}, respectively. Here the model's subcomponent deserves special attention: Naively treating the problem, the smaller clump's presence acts as a perturbation to the total system's phase-space density, and thus it is trivially in accordance with the estimated TG limit since the main clump is. However, this approach typically leads to overestimating the TG limit, considering that the secondary clump should be regarded as a bound object by itself. Taking the view that A2390 has undergone recent merger activity, it seems reasonable to assume that the subcomponent has formed at a sufficiently earlier time, and therefore it should be subject to its own phase-space distribution. This suggests that one should examine the secondary clump separately.

\begin{figure}
\includegraphics[trim = 40 0 0 0,width=0.85\linewidth]{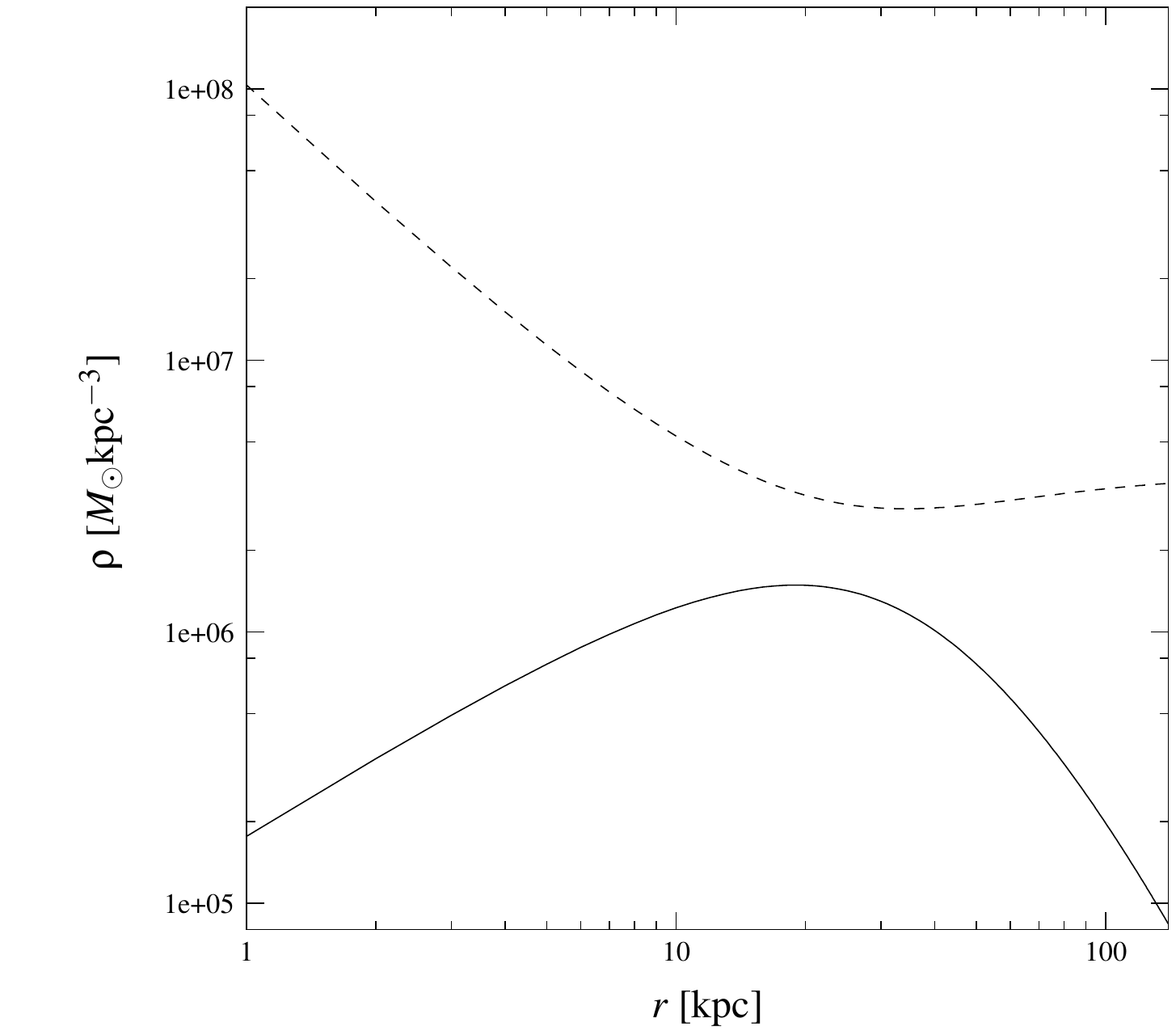}
\caption{Same as Fig. \ref{figure10}, but assuming $r_{C}\approx 10''$ ($37$kpc) and $\sigma_{\infty}\approx 440$km s$^{-1}$.}
\label{figure10a}
\end{figure}

Considering the subclump as an isolated object entirely dominated by $11$eV SNs, we have repeated the above TG analysis for the range of parameters listed in Ref. \cite{Pierre1996} (see Table \ref{table2}). To achieve a rather realistic TeVeS mass estimate, we have adopted a free interpolating function corresponding to Eq. \eqref{eq:22} for our calculations. The obtained SN density profile and the TG limit according to Eq. \eqref{eq:22a} are illustrated for two cases in Figs. \ref{figure10} and \ref{figure10a}. Assuming $r_{C}\approx 7''$ ($26$kpc) and $\sigma_{\infty}\approx 500$km s$^{-1}$, the subcomponent's density slightly exceeds the TG limit (up to $30$\%) within a range of approximately $10-25$kpc. Moving toward larger radii ($r\gtrsim 50$kpc), the SN density consistently stays below this limit. For a less compact model with $r_{C}\approx 10''$ ($37$kpc) and $\sigma_{\infty}\approx 440$km s$^{-1}$, the TG bound is never exceeded. Generally, our results seem to rule out configurations where the subclump is modeled with small values of $r_{C}$ ($\lesssim 8-9''$) whereas the bimodal TeVeS lens appears consistent with $11$eV SN HDM for larger choices of the core radius. Before drawing such a conclusion, however, we need to consider how strong the implication of the present analysis really is.

\begin{figure}
\includegraphics[trim = 40 0 0 0,width=0.85\linewidth]{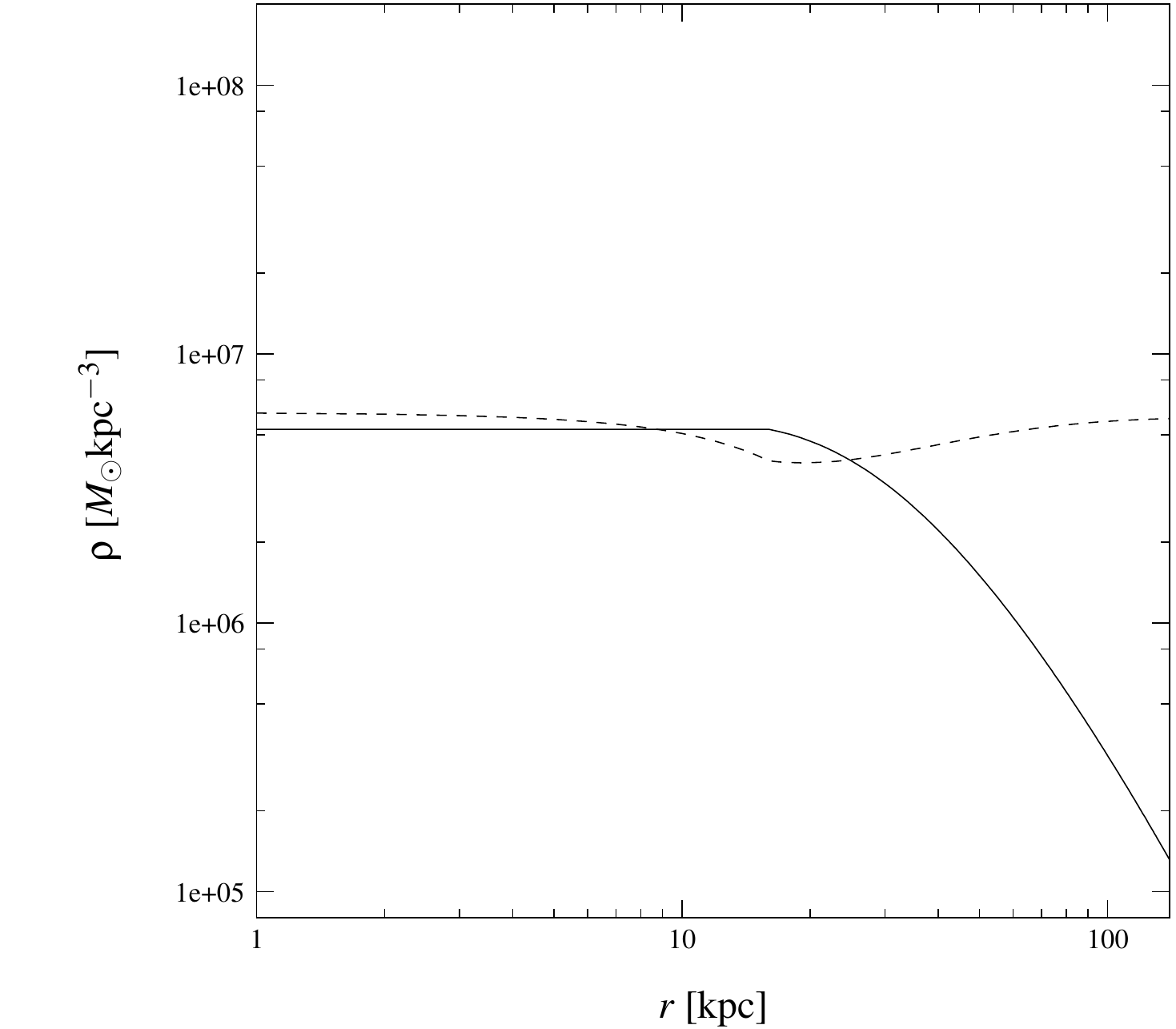}
\caption{Same as Fig. \ref{figure10}, but now assuming a uniform core of the SN distribution. Note that the radius at which the density becomes constant ($r\approx 15$kpc) is fixed by requiring a continuous distribution of SNs.}
\label{figure10b}
\end{figure}

First of all, we note that the lens model is based on the PIEMD model given by Eq. \eqref{eq:39}. This clearly introduces a bias on our estimates; other assumptions about the components' density distributions might yield a different result. In particular, the PIEMD model leads to an unphysical drop of the central density which could affect our estimate of the TG bound. To check this, we have modified the central SN density profile of the subclump model presented in Fig. \ref{figure10} in favor of a uniform core, but without changing its properties beyond $r\approx 15$kpc (the arc appears at $r\approx 26$ from the subcomponent's center). The resulting density profile and the corresponding TG limit are shown in Fig. \ref{figure10b}. While the TG bound is still violated within $\sim 10-25$kpc, we see that the density limit is notably decreased in the center, almost matching the assumed SN distribution. Therefore, it is unlikely that shifting matter to the central region can help to avoid an excess of the TG bound. Next, our estimates assume that the subclump can be treated as an isolated object. Since the clump resides within the background field of the main component, this is not rigorously true. Using Eq. \eqref{eq:42}, we find that the main component provides an external Newtonian gravitational field of around $a_{0}$ at the subclump's position. As for the subcomponent, this modifies the relation between gravitational field and underlying density distribution, and gives rise to an increase of the central SN density on the order of unity. Such a density boost could push seemingly consistent subclump models with $r_C\gtrsim 10''$ toward or even beyond the TG limit, but detailed statements about this issue are very sensitive to the actual model parameters.

Another point is related to the fact that our calculations rely on completely SN dominated lens components within $\sim 100$kpc. If placed at or close to the subclump's center, already a relatively small, concentrated baryonic mass, e.g. a galaxy, on the order $10^{8}-10^{9}M_{\odot}$ could help to relax the density constraint due to the TG limit \footnote{G. W. Angus (private communication).}. Whether such an approach can be reconciled with observations of this region, however, remains to be seen. Last but not least, we also need to check the viability of the current estimate of the TG limit which has been derived under simplified conditions. In what follows, we shall discuss in more detail how these simplifications affect our analysis.

As previously mentioned, the strong lensing domain in the center of A2390 is not in equilibrium and has a rather complicated nonsymmetric density distribution. This will obviously have an influence on the estimates for the TG limit. Considering nonequilibrium configurations in general, the velocity dispersion $\sigma$ is expected to increase for a given matter distribution when moving away from equilibrium. Since the value of $\sigma$ increases in this case, one would also obtain a higher density limit for SNs according to Eq. \eqref{eq:22a}. Taking the additional asymmetry into account, however, the situation becomes less clear. Depending on the system's actual properties, the TG limit could increase or decrease, and there seems to be no universal rule allowing one to make solid statements for anisotropic systems. Finally, one should also adhere to deviations from a Maxwellian velocity distribution. This issue has been addressed in Ref. \cite{Kull1996}. There it has been shown that the actual physical density limit becomes larger than the previous estimates of the TG limit which can be exceeded by up to a factor of $2$. Again, this would imply that SNs could account for more mass in the cluster. Combining the three aspects from above, it seems reasonable to assume that the true density limit will be on average higher than our previous estimates, meaning that density models with SNs become more flexible. Note, however, that such an argument generally does not replace the need for a rigorous treatment of particular systems.

Given the accuracy of our present analysis and accounting for all of the above, we conclude that the bimodal TeVeS lens model for the arc is in accordance with the assumption of $11$eV SNs. Nevertheless, it seems intriguing that the needed amount and distribution of $11$eV SNs lies so close to what they can maximally contribute to the system.
It should be obvious that all of our statements depend on the assumed lens configuration and are valid only for the bimodal model we have considered here. In particular, the bimodal lens model ignores the contribution of galaxies. These can have a significant impact on the lensing maps (see Sec. \ref{section521}), which is especially true for the galaxy $2592$ adjacent to the straight arc. A more realistic approach including all visible components would be useful to further constrain the properties of additional substructure and check whether such configurations remain consistent with respect to the TG limit. While our analysis is concerned only with the straight arc, the cluster A2390 actually exhibits a number of lensing features which should all be taken into account for a complete cluster model. Extending the investigation also to other massive galaxy clusters, future work should address such complex lens models and their implications for TeVeS or related theories and $11$eV SNs; a systematic way for approaching this problem has been outlined above.

\section{Conclusions}
\label{section6}
In this work, we have suggested the use of strong gravitational lensing by galaxy clusters as a test of the combined framework of TeVeS and massive SNs. Originally motivated by theoretical and recently also experimental particle physics \cite{Shaevitz2008,maltoni2,Giunti2008}, the idea of SNs with a mass around $11$eV has gained further interest as it provides a possible remedy for the problems of TeVeS and related theories ranging from large cosmological scales down to galaxy clusters. Unlike conventional CDM, such a fermionic HDM component is subject to strong phase-space constraints imposed by the TG limit. This allows one to check cluster lens models inferred within the above framework (or related ones) for consistency.

As an example, we have studied the cluster lens A2390 with its notorious straight arc. Because of its elongation and orientation, the straight image appears to be quite unusual and indicates the need for a rather special lens configuration. Adopting the approximation for weak fields and quasistatic systems, one of the main problems associated with the lensing analysis is the nonlinear relation between the TeVeS metric potential and the underlying matter density distribution. This nonlinearity prevents one from working with projected quantities and requires one to perform all calculations in three dimensions. In addition, one is left with a nontrivial, Poisson-type partial differential equation for the TeVeS scalar field.

To make some progress, we have considered a class of cluster models, based on the assumption of hydrostatic equilibrium, and investigated their lensing properties. This has been achieved by employing a MPI parallel solver for the TeVeS scalar field equation and simulating the corresponding lensing maps on the HUYGENS supercomputer which is located in Amsterdam. Our results imply that such quasiequilibrium configurations are not capable of explaining the observed straight arc. In particular, we have found no evidence for the formation of beak-to-beak or lips catastrophes \cite{a2390straight} due to intrinsic TeVeS effects, which could give rise to straight images. Line-of-sight effects and the impact of perturbations are typically small, changing the quantities of interest only on the order of a few percent. Similar to the situation in GR, a suitable TeVeS lens model therefore needs substantially more mass as well as a special density distribution in the cluster's core region.

Based on the above results, we have further outlined a general and systematic approach to cluster lenses which significantly reduces the problem's complexity by avoiding the need of solving the TeVeS scalar field equation. Combined with conventional lensing tools, this opens a new window to strong gravitational lensing in TeVeS-like modified gravity theories. As a first application, we have explored the TeVeS analog of the bimodal lens configuration discussed in Ref. \cite{Pierre1996}. For this model, we have derived the SN distribution necessary to produce the desired image, using a simplified approach. The obtained SN density profile has then been compared to the maximally allowed contribution set by the TG phase-space constraint. To this end, we have estimated the maximal density due to the TG limit following the prescription of Ref. \cite{sncluster} and found a slight excess of this limit for the model's secondary component if its core radius is small ($r_C\lesssim 8''-9''$). For less compact models, however, the TG bound is not violated. Given the accuracy of our current analysis, we therefore conclude that the bimodal TeVeS lens model appears consistent with the hypothesis of $11$eV SNs.

Note that the bimodal lens model ignores the contribution of galaxies. As has been shown, these can have a significant impact on the lensing maps. A more realistic approach, including all visible components and other lensing constraints, should be taken into account to obtain better bounds on the required SN distribution and to check whether such configurations remain consistent with respect to the TG limit. Future work should address more accurate ways of estimating the TG limit in this context, and we suggest extending the investigation to other massive galaxy clusters which indicate the need for dark substructure. Unless one considers different solutions to the missing mass problem inherent to this particular kind of modifications (see, e.g., Ref. \cite{Li2009}), the basic approach presented here should apply to any class of tensor-vector or tensor-vector-scalar theory which recovers the dynamics of MOND in the nonrelativistic limit. Lensing by galaxy clusters could therefore provide an interesting discriminator between CDM and such modified gravity scenarios supplemented by SNs. In addition to the above, we note that next-generation neutrino experiments \cite{Calligarich2010,Suzuki2010,Zito2010} will further constrain the plausibility of $11$eV SNs. Even if they remain viable candidates, it still needs to be seen whether such SNs do actually cluster in the desired way \cite{sncluster}.

Finally, we advert to the fact that our analysis neglects possible contributions due to perturbations of the TeVeS vector field $U_{\mu}$. Such contributions are known to be crucial for the formation of large-scale structure \cite{dod1,tevesneutrinocosmo}, where they provide the key to enhanced growth while perturbations of the scalar $\phi$ only play a subordinate role. As already pointed out in the literature \cite{tevesreview}, this typically affects scales $\gtrsim 0.1-1$Mpc and could be important for galaxy clusters. Owing to the more sophisticated structure of the field equations, however, even a rough magnitude of the vector's impact on these scales has not been estimated yet. Thus our work emphasizes the need for a quantitative description of these vector instabilities on small to intermediate scales, i.e. $\sim 0.01-1$Mpc. We also note that the result of such an analysis could strongly depend on the particularly assumed theory.

Despite the previously mentioned limitations of the present work, our numerical simulations are probably by far the most detailed in the context of TeVeS and certainly provide the first extensive study of strong lensing features within this modified gravity framework. Applications of our grid-based lensing code  (e.g. with respect to offsets between visible matter and weak or strong lensing features \cite{Shan2010,Zhaoclusterlens}) hold the promise of very constraining limits on TeVeS(-like) theory combined with HDM and other unified recipes for the dynamics of MOND and DM \cite{Zhao2010,Li2009,Milgrom2009a,Milgrom2010}.

\begin{acknowledgments}
The work has been performed under the HPC-EUROPA2 project (Project No. 228398) with the support of the European Commission - Capacities
Area - Research Infrastructures. M.F. thanks Garry Angus, Benoit Famaey and Matthias Bartelmann for valuable comments and discussions. Also, M.F. is deeply indebted to Garry Angus for sharing results and routines related to equilibrium distributions of massive neutrinos in A2390. J.L.G.P. thanks the cosmology group at the Institute of Theoretical Astrophysics in Heidelberg for their warm hospitality during the summer of $2007$ when this project came to life. C.F. acknowledges financial contribution from Contracts No. ASI-INAF I/023/05/0, No. ASI-INAF 088/06/0, and No. ASI ``EUCLID-DUNE'' I/064/08/0. H.S.Z. acknowledges partial support from the Dutch NWO funding through H.H. and support from the SARA supercomputing facility in Amsterdam. M.F. is supported by the Scottish Universities Physics Alliance (SUPA).
\end{acknowledgments}

\appendix

\section{Modeling the baryonic content of A2390}
\label{appendix1}
\subsection{X-ray gas and central mass distribution}
\label{appendix11}
To derive a reasonable model for the gas distribution in A2390, we use the results given in Ref. \cite{Allen2001}. The intrinsic electron density derived from CHANDRA observations (shown in Fig. $10$ of Ref. \cite{Allen2001}) can be well described by a spherical profile of the following form:
\begin{equation}
n_{e}(r) = \frac{n_{0}}{\left (1+\left (r/r_{0} \right )^{2} \right )^{1/2}},
\label{eq:18}
\end{equation}
where $n_{0} = 0.1$cm$^{-3}$ and $r_{0} = 10$kpc. Assuming a mean molecular weight of $w=0.6$ and an additional factor of $1.2$ to account for the global effect of the cluster's stellar components, we thus obtain an expression for the effective central density profile with a central density of $\rho_{0} = 1.8 \times 10^{6}M_{\odot}$ kpc$^{-3}$. Since the volume integral of Eq. \eqref{eq:18} diverges, we smoothly cut the profile at radius $R$ within a range of $200$kpc. The cutoff scale is set to $R=1$Mpc which corresponds to $0.7 r_{500}$ \footnote{Assuming the framework of GR with CDM, the overdensity radius $r_{500}$ is the radius within which the mean matter density is $500$ times the critical density of the universe at the cluster's redshift.} as given in Ref. \cite{Vikhlinin2006}. This yields a total integrated mass of $M\sim 1.3\times 10^{14} M_{\odot}$ and a surface density profile which is in good agreement with a $10-20$\% gas fraction of the enclosed projected lensing mass estimated in the framework of GR \cite{Squires1996,Hoekstra2007}. The density distribution specified by Eq. \eqref{eq:18} is illustrated in Fig. \ref{figure1} (dotted line).

\begin{figure}
\includegraphics[trim = 80 0 0 0,width=0.9\linewidth]{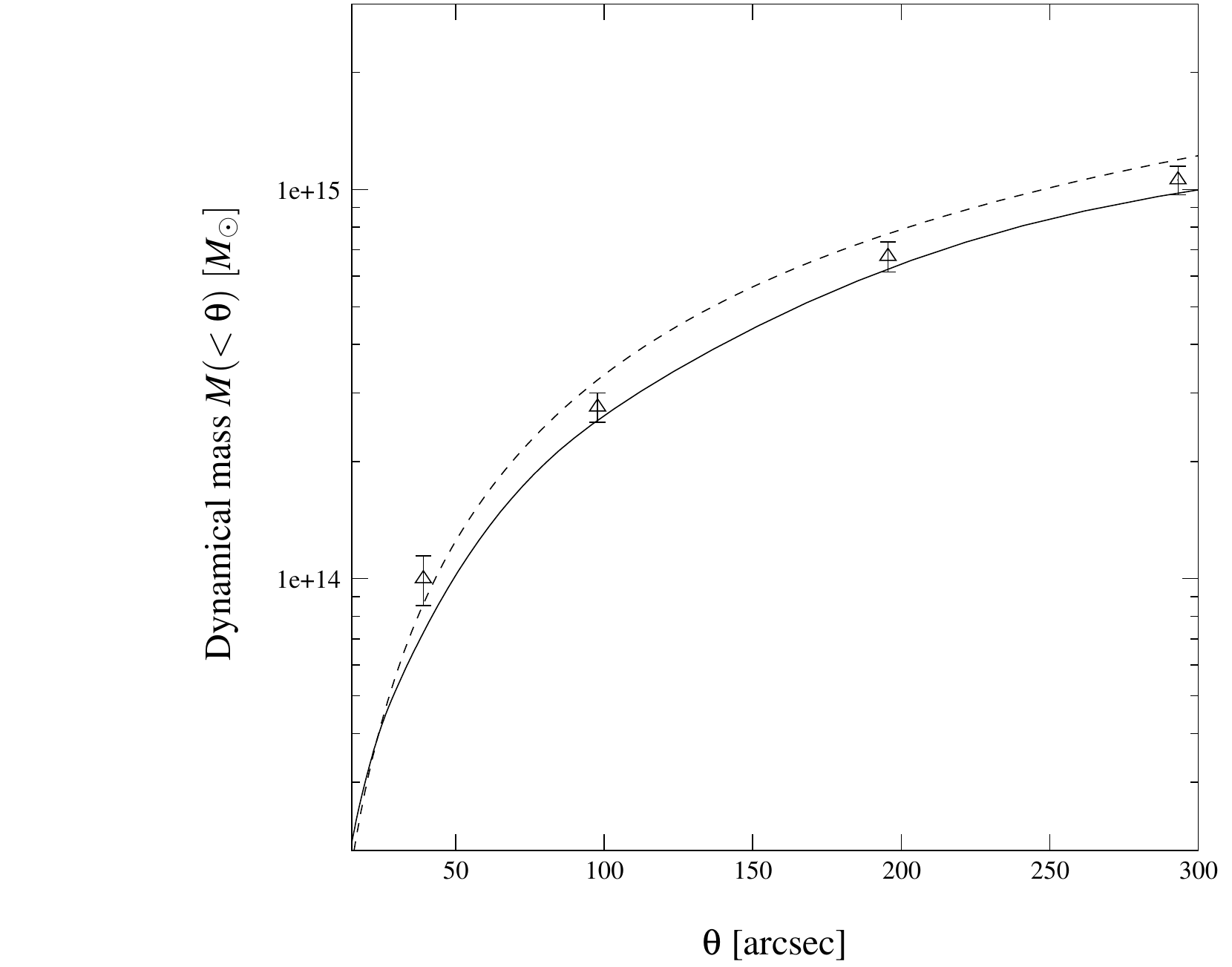}
\caption{Enclosed projected (Newtonian) dynamical mass profiles for our TeVeS equilibrium model (gas + SNs; solid line) and an NFW model (dashed line). The lower mass of the TeVeS model is mostly caused by the approximate description of the gas density given in Eq. \eqref{eq:18}; triangles indicate the estimates from weak lensing observations. At the arc's position ($\theta\approx 38''$), the relative difference between the models is about $10$\%.
}
\label{figure2}
\end{figure}

Although our choice for the density profile is less accurate and results in a slightly smaller mass than typical $\beta$ models \cite{Pierre1996,Allen2001} or more flexible ones \cite{Vikhlinin2006}, it will be sufficient for our analysis. As is shown in Sec. \ref{section43}, the relevant lensing mass is mostly dominated by the contribution of SNs. Thus the strong lensing results, which we are primarily interested in here, will be relatively insensitive to the actual assumption of the central baryonic distribution. Adopting the more realistic density models above in a few selected simulation runs, otherwise identical to those presented in Sec. \ref{section52}, we find only small differences on the order of a few percent in the corresponding results and confirm our argument. This is also indicated by comparing the enclosed projected dynamical mass profiles of our cluster model (gas + SNs) to the Navarro-Frenk-White (NFW) profile \cite{NFW} estimated in Ref. \cite{Vikhlinin2006} (see Fig. \ref{figure2}). Although the TeVeS model underestimates the mass, the discrepancy from the NFW model is only about $10$\% at the arc's position ($\theta\approx 38''$). In addition, the figure shows the weak lensing results obtained from the Canada-France-Hawaii Telescope (CFHT) for a photometric redshift distribution based on the CFHT Legacy Survey data \cite{Hoekstra2007,Mahdavi2008}. The relative good agreement between dynamical and weak lensing mass estimates further implies that structure along the line of sight plays no significant role and does not affect our analysis. All presented quantities have been corrected for the cosmological model specified in Sec. \ref{section22}. Note, however, that a rather accurate description of the gas density as well as its temperature profile is important to estimate the neutrino content necessary for hydrostatic equilibrium in TeVeS \cite{sncluster}.

\subsection{Galaxy morphology and masses}
\label{appendix12}
Since a rather detailed model of the cluster might be important in TeVeS, we also need to take the contribution of individual galaxies into account. For simplicity, we only consider the most massive member galaxies in the immediate vicinity of the straight arc's observed position; galaxies which are located farther away are unlikely to affect the TeVeS lensing maps in this region, which is confirmed by our results presented in Sec. \ref{section52}. Although A2390 exhibits a rich class of galaxy morphologies, with many galaxies showing elliptical or lenticular shapes, the impact of individual morphologies on the arc's environment can safely be neglected due to the galaxies' sufficiently large distances. While this is not necessarily true for the galaxy $2592$ (see Fig. \ref{figure0}) which resides directly adjacent to the arc, a spherical density model provides a good description, which is indicated by the rather mild ellipticity seen in the optical HST image. As can be seen in Sec. \ref{section52}, this approximation does not affect the basic results of our analysis - at least in the case of quasiequilibrium configurations.

Furthermore, we assume that all considered galaxies can be modeled by a matter distribution of the form
\begin{equation}
\rho(r) = \frac{Mr_{H}}{2\pi (r+\epsilon)(r+r_{H})^{3}},
\label{eq:19}
\end{equation}
where $\rho(0)=M/(2\pi\epsilon r_{H}^{2})$ is the central matter density, and the profile's core radius is universally set to $r_{H}=3$kpc. The length scale $\epsilon$ corresponds to a smoothing parameter becoming necessary due to the limited resolution of our simulations and is specified in App. \ref{appendix22}. For $\epsilon =0$, Eq. \eqref{eq:19} reduces to the well-known Hernquist profile \cite{hernquist} which closely approximates the de Vaucouleurs $R^{1/4}$ law for elliptical galaxies.

To infer the masses of individual galaxies, needed for our strong lensing analysis, we consider the data of the spectro-photometric catalog compiled in Ref. \cite{Fritz2005}, which lists magnitudes for $48$ galaxies inside the cluster A2390. All magnitudes are given in the Gunn $r$ band \cite{Thuan1976}, and a simple formula \cite{Moro2000} to convert the $R$ Johnson magnitude and the $B-V$ color index to the Gunn $r$ band can be found in the literature \footnote{An excellent description of the Gunn magnitude system is given on the website \protect\url{http://ulisse.pd.astro.it/Astro/ADPS/}.}. Accordingly, we have computed $r_\odot$, the Gunn $r$ magnitude of the sun, adopting $R_\odot=4.42$ \cite{BiMe1998} and $(B-V)_\odot = 0.64$ \cite{Holmberg2006}. We have found $r_\odot = 4.95$ which is rather close to the $r$ value inferred from SDSS, the corresponding band being quite similar to the Gunn $r$ band. Our result for $r_\odot$ has then been used to evaluate the absolute luminosities of the galaxies given in Ref. \cite{Fritz2005}.

Next, we need a realistic mass-to-light ratio ($M/L$) in order to determine the galaxy masses. To this end, we have followed a twofold approach: First, we have adopted a constant $M/L$ derived by combining the relation between $M/L$ and the $g-r$ color index presented in Ref. \cite{Bell2003} with the $g-r$ colors for massive ellipticals in the red sequence of the SDSS given in Ref. \cite{Blanton2005}. The corresponding masses are labeled as $M_{1}$. Second, we have also considered $M/L$ as a function of $M$ in agreement with the results for the galaxies of A2390 discussed in Ref. \cite{Fritz2005}. For this, a dynamical mass estimate based on measured velocity dispersions was used. As elliptical galaxies are mostly subject to the strong gravity regime within their half-light radius, however, estimates in both MOND/TeVeS and Newtonian dynamics should be roughly the same. This second mass estimate, denoted as $M_{2}$, is probably more reliable since it involves fewer assumptions. The properties of the such obtained galaxy models are listed in Table \ref{table1}.

\subsection{Role of the central cD galaxy}
\label{appendix13}

Assuming an equilibrium model for A2390, it has been found that $11$eV SNs reach their densest possible configuration for $r\lesssim 20$kpc (see Sec. \ref{section43}). Since Eq. \eqref{eq:25} takes the TG bound into account, our cluster model misses some mass in the central part and does not correspond to a genuine equilibrium situation. A way of compensating for this is to consider an additional contribution due to the central cD galaxy. Following the lines of Ref. \cite{sncluster}, one can estimate a total galaxy mass of approximately $M=1.8\times 10^{12}M_{\odot}$. As the central region of A2390 is neither spherically symmetric nor in equilibrium \cite{Allen2001,Vikhlinin2006}, it is important to note that such an approach has no real physical meaning, but rather offers a convenient way to tweak our cluster model.

What does the above mean for our lensing analysis? Modeling the cD galaxy as a point mass, a straightforward calculation shows that its impact on the TeVeS lensing maps can be safely neglected. At the position of the straight arc ($38''$ or $140$kpc from the cluster center), the additional matter gives rise to changes of $1-2$\%. Moving to smaller radii, the deviation grows, but we are not interested in this region anyway. Thus we consider the cluster model presented in Sec. \ref{section4} as sufficient for our investigation.

\section{Numerical tools and setup}
\label{appendix2}
\subsection{Solving the scalar field equation}
\label{appendix21}
Having set the framework of gravitational lensing and cosmology in Sec. \ref{section2}, we may proceed with calculating the desired TeVeS lensing maps. The main problem associated with this task is to solve the scalar field equation specified in Eq. \eqref{eq:5} which can be rewritten as
\begin{equation}
\Delta\phi = \bar\rho,
\label{eq:26}
\end{equation}
where the effective density $\bar{\rho}(\rho ,\partial_{i}\phi , \partial_{i}\partial_{j}\phi)$ is
\begin{equation}
\bar\rho = \frac{kG}{\mu}\rho-2\frac{kl^{2}}{\mu}\frac{\partial\mu}{\partial y}\left((\partial_{i}\phi)(\partial_{j}\phi)(\partial_{i}\partial_{j}\phi)\right),
\label{eq:27}
\end{equation}
and indices run from $1$ to $3$. Equation \eqref{eq:26} corresponds to a nonlinear second order elliptic boundary value problem and can be tackled numerically. A Fourier-based solver employing an iterative relaxation scheme and an equidistant grid has been presented in Paper I where the basic algorithm and involved approximations are extensively discussed.

One of the numerical challenges of our analysis is that we need to resolve galactic scales in a cluster-wide box, which requires a relatively large number of grid points. Since all calculations have to be performed in three dimensions, this clearly exceeds the capacity of a single-processor machine, in terms of both needed time and memory, and therefore calls for a more powerful computer architecture. For this reason, we have implemented a parallel version of the original solver using the Message Passing Interface (MPI) standard. The parallelization as well as all calculations presented in this work have been carried out on the HUYGENS supercomputer at SARA in Amsterdam within the HPC-EUROPA Transnational Access Programme. The HUYGENS system consists of $104$ nodes, with $16$ dual core processors (IBM Power6, 4.7 GHz) as well as either 128 GBytes or 256 GBytes of memory per node, thus providing an excellent environment for our needs.

\begin{figure}
\includegraphics[trim= 30 0 0 0,width=0.9\linewidth]{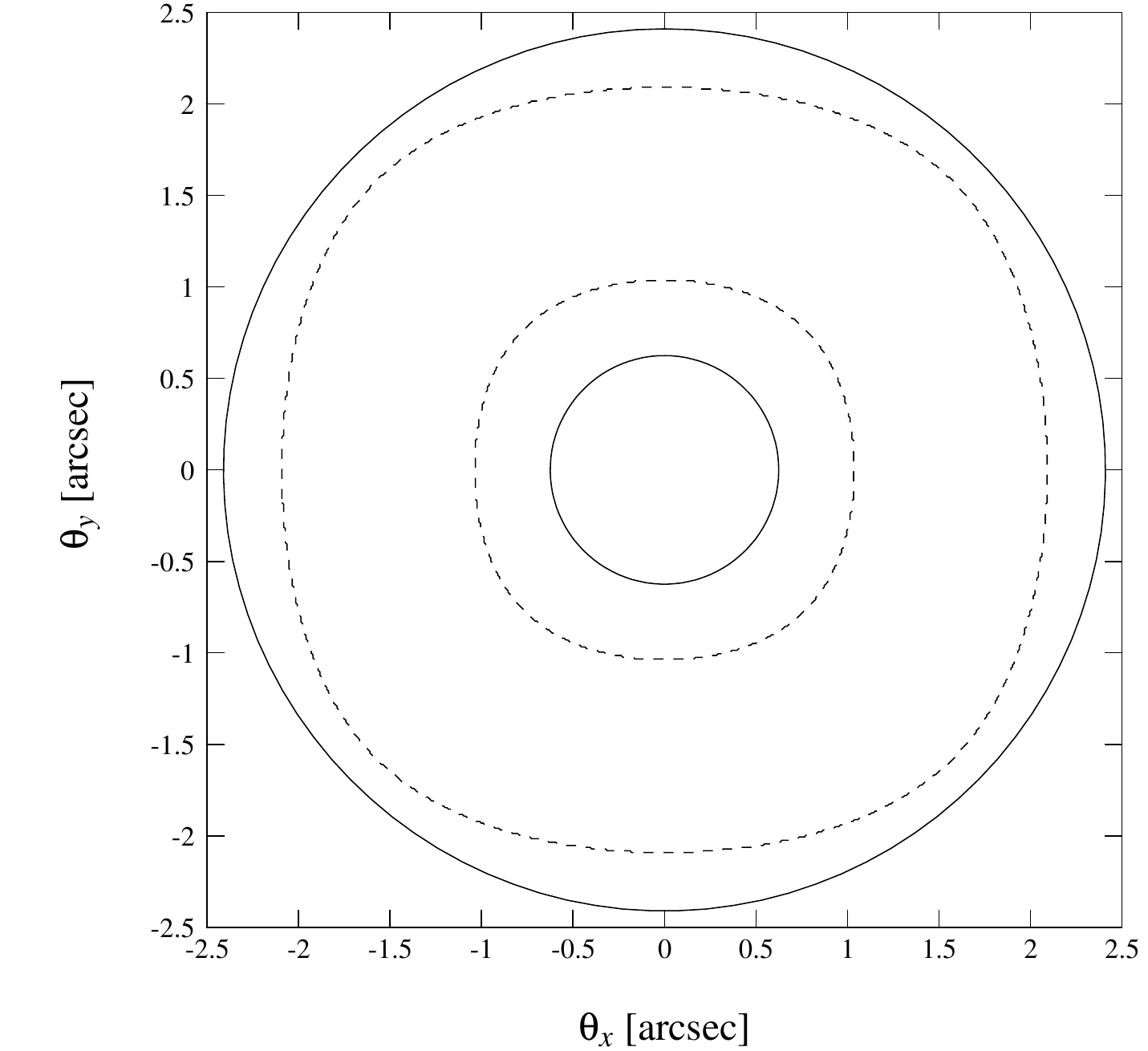}
\caption{Predicted critical curves for an isolated galaxy given by Eq. \eqref{eq:19}: Assuming an aperture mass of $3.5\times 10^{11}M_{\odot}$ within a $3''$ ($\sim 11$kpc) diameter as well as a lens and source redshift of $z_{l}=0.23$ and $z_{s}=1$, respectively, we present results for both a high resolution ($\sim 0.05''$, solid line) and a low resolution setting ($\sim 1.2''$, dashed line) with subsequent interpolation.}
\label{figure3}
\end{figure}

The parallel solver has been tested with analytic TeVeS models such as the Hernquist lens (see, e.g., Ref. \cite{lenstest} or Paper I), and has also been compared to previous calculations for the ``bullet cluster'' (Paper I), yielding exactly the same results - up to machine accuracy - as the serial version for identical input parameters. Considering the numerical setup for A2390, we choose a physical box size of $V=d^{3}=(4$Mpc$)^{3}$ in order to meet the requirements of the point lens approximation at the grid's boundaries (for details see Paper I). Performing a variety of test runs, we have found that the solver's convergence property quickly deteriorates if we increase the number of grid points per dimension $N$, meaning that the code takes many iteration steps or even fails to converge \footnote{It is quite likely that the problem is partly related to the destabilizing influence of high frequency modes. These modes are able to ``see'' and amplify numerical artifacts which are present both in regions around local extrema, where the derivative of scalar potential exhibits values close to zero, and at the grid's boundaries.}. Typically, this problem already occurs at $N=512$ and manifests itself through extreme fine-tuning of the constant relaxation parameter $\omega$ which determines the scalar's mixing of at each iteration step,
\begin{equation}
\Delta\tilde{\phi}^{(n)} = \bar{\rho}^{(n)},\quad \phi^{(n+1)}=\omega\tilde{\phi}^{(n)}+(1-\omega)\phi^{(n)}.
\label{eq:28}
\end{equation}
Depending on the particularly used density model of the cluster, acceptable values for $\omega$ vary within a range of $0.7-0.9$, but allow them to be easily identified just after a few iterations. Compared to the analysis of Paper I, we thus obtain no universal value for the relaxation parameter. Similarly, we also note that the solver's behavior becomes more sensitive with respect to the scalar's initial guess. This is expected because the effective deviation from the desired solution increases with $N$, and can usually be accounted for by slightly modifying the original point mass ansatz of Paper I to achieve a finite core,
\begin{equation}
\phi^{(0)}(r) \propto \log(r+r_{c}),
\label{eq:29}
\end{equation}
where $r_{c}$ is on the order of a few $d/N$. While more elaborated guesses are also possible, they typically do not yield a much better performance.

\subsection{Numerical setup for A2390}
\label{appendix22}

In all simulation runs, we set the number of grid points per dimension to $N=896$. This yields a resolution of approximately  $1.2''$ ($\sim 4.5$kpc) for our choice of $d=4$Mpc. To improve the numerical stability of our Fourier solver, we further require all density components to be centered within their respective subcube, which can lead to a maximal deviation of $0.6''$ from the positions listed in Table \ref{table1}. In addition, we assume a smoothing parameter $\epsilon = 1$kpc for the galaxy profile given by Eq. \eqref{eq:19}. Once the desired fields and derivatives are calculated, we use a cubic spline to interpolate our results and determine the relevant lensing quantities. For the given specifications, individual simulation runs typically require $30-50$ iteration steps to converge, and can last up to $24$ hours using $32$ processors.

The interpolation approach is justified because the exact result is expected to be relatively smooth. To support this argument, we performed a small numerical experiment: Assuming an aperture mass of $3.5\times 10^{11}M_{\odot}$ within a $3''$ ($\sim 11$kpc) diameter and the parameters from above, we compared the predicted critical curves of an isolated galaxy given by Eq. \eqref{eq:19} for low resolution ($\sim 1.2''$) with subsequent interpolation to those calculated for a higher resolution setting ($\sim 0.05''$). Choosing a lens and source redshift of $z_{l}=0.23$ and $z_{s}=1$, respectively, the results are shown in Fig. \ref{figure3}. While the radial critical curve is not very well recovered, the radius of the tangential critical curve, which is relevant for our considerations on the straight arc \footnote{Although radial caustics can produce straight images, the resulting orientation (pointing towards the center of the corresponding lens) is not compatible with the observed arc.}, is only underestimated by roughly $10$\% on average. Considering the full cluster model of A2390, however, galaxies are not isolated, but reside within the cluster's background field, which leads to a boost of their corresponding Einstein radii. Therefore, we expect the accuracy of the calculated lensing properties, including critical curves and caustics, to be significantly improved and sufficient for our analysis in this case.

\bibliography{ref_short}

\end{document}